\documentclass[sigconf, 9pt, nonacm]{acmart}

\usepackage{algorithmic}
\usepackage{graphicx}
\usepackage{textcomp}
\usepackage{xcolor}
\usepackage{cleveref}
\usepackage{tikz}
\usetikzlibrary{shapes.symbols, positioning, arrows.meta, calc}
\usepackage[T1]{fontenc}
\usepackage{url}
\usepackage{xspace}
\usepackage[printonlyused]{acronym}
\usepackage[most]{tcolorbox}
\tcbuselibrary{minted, skins, listings, breakable}
\usepackage{minted}
\usepackage{fancyvrb}
\renewenvironment{minted}[2][]{%
    \VerbatimEnvironment
    \footnotesize
    \begin{Verbatim}[breaklines=true,breakanywhere=true]%
}{%
    \end{Verbatim}%
}
\usepackage{fontawesome5} 
\usepackage{makecell} 
\usepackage{tabularx} 

\usepackage{subcaption}
\newcommand{\proj}{\textsc{PatchyBFT}\xspace}
\newcommand{\magic}{\textsc{PatchyBFT}\xspace}

\newcommand{\locRemovedPercent}{65\%\xspace}
\newcommand{\maxUnique}{7\xspace}

\usepackage{pgfplots}
\usepgfplotslibrary{ternary}
\pgfplotsset{width=10cm, compat=1.18}

\usepackage[colorinlistoftodos,prependcaption,textsize=tiny]{todonotes}

\newcommand{\para}[2]{
    \begin{tcolorbox}[
        title=#1,        
        colback=blue!5,         
        colframe=white!75!black, 
        fonttitle=\bfseries,    
        width=\linewidth,       
        sharp corners,          
        boxrule=0.5mm,          
        coltitle=black          
        ]
#2
\end{tcolorbox}
}
\renewcommand{\para}[2]{
#2
}

\newcommand{\answer}[3][]{
    \begin{tcolorbox}[
        title={\ifx #1 \empty #2 \else #2 \hfill| #1 \fi},  
        colback=blue!1,         
        colframe=white!90!black,
        fonttitle=\bfseries,    
        width=\linewidth,       
        sharp corners,          
        boxrule=0.5mm,          
        boxsep=0.1mm,           
        left=1.0mm,             
        right=1.0mm,            
        top=1.0mm,
        bottom=1.0mm,
        coltitle=black          
        ]
#3
\end{tcolorbox}
}

\newtcolorbox{humanbox}{%
    enhanced,
    colback=humanbg, colframe=humanbg,
    boxrule=0pt, arc=2pt,
    left=4pt, right=4pt, top=3pt, bottom=3pt,
    before skip=3pt, after skip=3pt,
    overlay={%
        \node[anchor=north east,
              font=\bfseries\footnotesize,
              text=humanlabel,
              inner sep=0pt]
            at ([xshift=-6pt, yshift=-3pt]frame.north east) {Human fix};
    }
}

\newtcolorbox{llmbox}{%
    enhanced,
    colback=llmbg, colframe=llmbg,
    boxrule=0pt, arc=2pt,
    left=4pt, right=4pt, top=3pt, bottom=3pt,
    before skip=3pt, after skip=3pt,
    overlay={%
        \node[anchor=north east,
              font=\bfseries\footnotesize,
              text=llmlabel,
              inner sep=0pt]
            at ([xshift=-6pt, yshift=-3pt]frame.north east) {LLM fix};
    }
}

\newtcolorbox{codebox}[1][]{%
    enhanced,
    colback=gray!4, colframe=black!65,
    boxrule=0pt, arc=2pt,
    left=4pt, right=4pt, bottom=3pt,
    top=\notblank{#1}{3pt}{3pt},
    before skip=3pt, after skip=3pt,
    overlay={%
        \notblank{#1}{%
            \node[anchor=north east,
                  font=\bfseries\footnotesize,
                  text=llmlabel,
                  inner sep=0pt]
                at ([xshift=-6pt, yshift=-6pt]frame.north east) {#1};%
        }{}%
    }
}
\tcbset{
  promptbox/.style={
    enhanced,
    listing only,
    colback=gray!4,
    colframe=black!65,
    boxrule=0.5pt,
    arc=1.5mm,
    left=1.5mm,
    right=1.5mm,
    top=1mm,
    bottom=1mm,
    fonttitle=\bfseries,
    coltitle=black,
    segmentation hidden,
lower separated=false,
    colbacktitle=gray!15,
    attach boxed title to top left={xshift=1.5mm,yshift=-1.8mm},
    boxed title style={
      boxrule=0.5pt,
      arc=1mm,
      colframe=black!65,
      colback=gray!15
    },
    listing options={
      basicstyle=\ttfamily\scriptsize,
      breaklines=true,
      breakatwhitespace=false,
      columns=fullflexible,
      keepspaces=true,
      showstringspaces=false,
      segmentation hidden,
lower separated=false,
    }
  }
}

\definecolor{humanbg}{HTML}{EAF4EA}     
\definecolor{humanlabel}{HTML}{1B5E20}  
\definecolor{llmbg}{HTML}{EAEEF7}       
\definecolor{llmlabel}{HTML}{0D2A6B}    
\definecolor{promptbg}{gray}{0.96}
\definecolor{promptframe}{gray}{0.55}
\lstdefinestyle{promptstyle}{
  basicstyle=\ttfamily\tiny,
  backgroundcolor=\color{promptbg},
  frame=single,
  framerule=0.4pt,
  rulecolor=\color{promptframe},
  columns=fullflexible,
  keepspaces=true,
  showstringspaces=false,
  breaklines=true,
  breakatwhitespace=false,
  postbreak={},
  tabsize=2,
  xleftmargin=1mm,
  xrightmargin=1mm,
  aboveskip=0.6\baselineskip,
  belowskip=0.4\baselineskip,
  captionpos=b
}
\usepackage{caption}
\usepackage{subcaption}
\DeclareCaptionSubType*{listing}
\newminted[rustsnippet]{rust}{%
    fontsize=\footnotesize,
    linenos=false,
    framesep=2mm,
    frame=single,
    rulecolor=\color{rulec},
    breaklines=true,
    autogobble
}

\acrodef{BFT}{Byzantine Fault Tolerant}
\acrodef{CoT}{Chain-of-Thought}
\acrodef{CFT}{Crash Fault Tolerant}
\acrodef{LLM}{Large Language Model}
\acrodef{CWE}{Common Weakness Enumeration}
\acrodef{TEE}{Trusted Execution Environment}
\acrodef{AST}{Abstract Syntax Tree}
\acrodef{ELF}{Executable and Linkable Format}
\acrodef{LoC}{Lines of Code}

\AtBeginDocument{%
  }

\setcopyright{none}
\begin{document}

\title{\proj: Automating Diversification of Fault-Tolerant Systems using LLMs}

\author{Arne Vogel}
\email{arne.vogel@fau.de}
\orcid{0000-0002-5163-165X}
\affiliation{%
  \institution{FAU Erlangen-Nürnberg}
  \city{Erlangen}
  \country{Germany}
}

\author{Christian Berger}
\orcid{0000-0003-2754-9530}
\affiliation{%
  \institution{FAU Erlangen-Nürnberg}
  \city{Erlangen}
  \country{Germany}}
\email{berger@cs.fau.de}

\author{Rüdiger Kapitza}
\orcid{0000-0002-8116-7763}
\affiliation{%
  \institution{FAU Erlangen-Nürnberg}
  \city{Erlangen}
  \country{Germany}}
\email{ruediger.kapitza@fau.de}

\renewcommand{\shortauthors}{Vogel et al.}


\begin{abstract}
Fault-tolerant agreement protocols fail if replicas share a common flaw that simultaneously affects more replicas than the tolerable threshold.
Therefore replicas should ideally fail independently, which can be achieved through diversification. 
However, in practice, often the same protocol implementation is shared by all replicas which is not surprising given that the provision of multiple diverse implementations is difficult and highly laborious. 
This poses a major risk, as a shared protocol implementation is a prime candidate for common bugs due to its complexity.

With \proj, we demonstrate how, given a reference implementation, \acp{LLM} can be utilised for the automated and scalable generation of code that compiles, passes tests, and crucially differs semantically/binary-wise, that can replace code in the reference implementation, thereby significantly reducing diversification costs.
We demonstrate the feasibility of diversification of replication protocol implementations using \acp{LLM} by diversifying three implementations: PBFT, HotStuff, and Raft, showing how up to \locRemovedPercent of the codebase can be diversified.
\end{abstract}

\acresetall




\settopmatter{printacmref=false}
\renewcommand\footnotetextcopyrightpermission[1]{}

\maketitle

\section{Introduction}

Fault-tolerant agreement protocols have become increasingly crucial for scalable web services~\cite{wenbing2006bftws}, cloud infrastructures~\cite{corbertt2013spanner}, and, more recently, distributed ledger systems~\cite{vogel2023sok, bessani2020blockchain, neiheiser2021kauri, gupta2019blockchain}.
Depending on the situation, these protocols can be crash or Byzantine fault-tolerant (BFT) and are designed to handle $f$ faults in $n=2f+1$ or $n=3f+1$ replicas, respectively, while still providing correct results~\cite{castro1999practical}.
In practice, replicas must be fault-independent with respect to common failure modes, such as shared vulnerabilities in replica implementations, deployment in the same region, or operation by the same operator, to achieve this theoretical guarantee.
In this work, we focus on diversifying the implementations of agreement protocols as otherwise a shared bug in the system could easily violate the limit of $f$ faulty replicas.
One of the core methods for providing fault-independent software for replicas is N-version programming~\cite{avizienis85n-version}. 
It requires the implementation of multiple, diverse versions of a system based on a common specification, ideally by different development teams using different programming languages or development methodologies, to minimize the risk of common-mode errors.
As a result, N-version programming is considered prohibitively resource-intensive and is typically not applied to common IT services, even though their unavailability or corruption can lead to a poor user experience and significant revenue losses.
Since the seminal work on making Byzantine Fault Tolerance practical for everyday IT services~\cite{castro1999practical}, the question of how to achieve fault independence in practice without adopting full N-version programming has remained an open topic especially relevant today with distributed ledgers managing billions in value~\cite{vogel2023sok, wang2022bft, rubambiza2023comosum}.
One line of work is \emph{opportunistic} N-version programming, which builds on software heterogeneity for well-established APIs~\cite{castro03base}. 
As an example, diversification at the operating system level is a direction that has been proposed~\cite{sousa2008forever, garcia2019lazarus}, as many systems provide a POSIX-compliant system interface so that the protocol implementation and the replicated application can run on a diverse set of operating systems that share almost no bugs.
Opportunistic N-version programming has also been explored at the replicated application or service level. 
 Examples include using different relational databases, as SQL is a common standard with various implementations~\cite{vandiver2007hrdb, gahsi04dbs}.
Finally, more general diversification mechanisms can be applied, such as address space layout randomization~\cite{spengler2001aslr} or introducing diversification during compilation~\cite{jackson2011compiler} such as function inlining, outlining, splitting, control flow flattening, or system call mapping randomization among others~\cite{larsen2014sok, cohen1993operating}, which have been used in distributed systems such as Spire~\cite{babay2018spire}.
Despite all these previous works, diversification at the protocol level, which is at the heart of replicated systems, has largely been abandoned.
This is not surprising, as implementing a fault-tolerant agreement protocol involves highly concurrent code featuring complex communication logic and requires the correct use of various cryptographic methods. 
Thus, even providing a single correct implementation that offers good performance is already a significant challenge. 
As a result, achieving diversification at the level of the Byzantine fault-tolerant agreement protocol via N-version programming has so far only been achieved for very few widely used protocols such as Ethereum~\cite{clientdiversify2024, mainnet2024}.

\acp{LLM} currently change the way code is written in industry, with 80\% of professional developers already using them, 50\% even daily~\cite{stackoverflow2025}.
They are used for code understanding~\cite{nam2024understanding_llm}, to generate new code~\cite{mathews2024tdd_llm}, explore the design space of programs~\cite{zamfirescu2025design}, and even for N-version programming to combat compiler bugs (limited to pure functions without side effects, not suitable for most agreement protocol implementation functions)~\cite{ron2024galapagos}.

In this paper, we propose \magic, which enables the highly automated and scalable diversification of \ac{BFT} and \ac{CFT} agreement protocol implementations using \acp{LLM}.
This is achieved by using multiple \acp{LLM} to generate diversified implementations based on an original version. 
However, naive code generation is insufficient, so we must address three challenges:
\begin{enumerate}
    \item At what level of abstraction should code diversification occur and how much context should be provided to an \ac{LLM} to drive the generation?
    \item How can we ensure that the diversified code maintains functional equivalence?
    \item How can we validate that diversification actually achieves representational and binary difference rather than merely syntactic variation?
\end{enumerate}

To address these challenges, we designed and implemented \proj, which automates the diversification of Rust-based distributed protocols.    
Rust is a modern systems programming language widely adopted in the systems research community~\cite{ruesch2019themis, jitaowang2024bftdiagnosis, gkagol2019aleph, bearer2024espresso}.
Despite the security properties of Rust~\cite{rebert:2024:google, stoep:2024:google, white-house:2024:report}, diversification is still relevant for defending against implementation-level vulnerabilities that escape Rust's compile-time safety guarantees~\cite{base64bug, isahcbug, hassain2024counterexamples, meneely2025justuserust}. 

To highlight the benefits of \magic, we automatically diversified Themis~\cite{githubthemis} (implementing the PBFT algorithm~\cite{castro1999practical, ruesch2019themis}), hotstuff\_rs~\cite{githubhotstuff} (implementing the HotStuff algorithm~\cite{yin2019hotstuff}), and Openraft~\cite{databend2024openraft}, a \ac{CFT} replication protocol implementation. 

\textbf{Contributions.}
\magic makes diversification practical for \ac{BFT} and \ac{CFT} protocols.
We demonstrate how to use \acp{LLM} to generate protocol changes at function level based on reference implementations.
For validation of these changes, we introduce \textit{safeguards} for functional and diversification correctness.
As a proof of concept, we additionally highlight that \acp{LLM} can go beyond simple diversification and even fix bugs in \ac{BFT} protocols, without having been informed of the specific bug.
Further contributions include:
\begin{itemize}
    \item \textbf{Automated protocol diversification:} We demonstrate how to use \acp{LLM} to generate representation/binary different implementations of complex replication protocols, with up to \locRemovedPercent of the original code being automatically changed.
    \item \textbf{Functional validation framework:} We show how to validate the generated code for functional equivalence with unit tests and protocol conformance tests against the reference implementation, ensuring changes do not break the protocol.
    \item \textbf{Representational/binary validation framework:} We introduce a new methodology for how to ensure that diversified code is representationally and binary different, using clone detection and binary difference validation.
\end{itemize}

The rest of the paper is structured as follows:
In Section~\ref{sec:background} we provide background information and related work on \ac{BFT} systems, \acp{LLM} and Rust.
This is followed by an investigation of using \acp{LLM} to create patches in Section~\ref{sec:proof-of-concept}.
Next, we describe the design of \magic and describe how \magic fits into the lifecycle of distributed systems in Section~\ref{sec:design} followed by implementation details in Section~\ref{sec:implementation}.
We evaluate \magic in Section~\ref{sec:evaluation} and show future work in Section~\ref{sec:limitations}.
Finally, we conclude in Section~\ref{sec:conclusion}.

\section{Background \& Related Work}
\label{sec:background}

    \textbf{Diversification of Byzantine Fault-Tolerant protocols.} 
    \ac{BFT} protocols can tolerate up to $f$ faulty replicas in a system of $3f+1$ replicas~\cite{castro1999practical, distler2021system}.
    This is a theoretical limit that requires fault-independent implementations in practice.
    Otherwise, an attacker can exploit the same bug in all replicas, easily exceeding the theoretical threshold of $f$ faulty replicas.
    Similarly, \ac{CFT} protocols have a threshold of $2f+1$ where up to $f$ crashes can be tolerated.
    Testing techniques are proposed to limit faults, but are unable to give comprehensive guarantees~\cite{bano2022twins}, and while formal models~\cite{hawblitzel2015ironfleet, lesani2016chapar, wilcox2015verdi} can give comprehensive correctness guarantees, they are laborious (3.7 person years reported)~\cite{hawblitzel2015ironfleet} and can still contain bugs through assumptions made for the formal specification~\cite{fonseca2017formal_emerical}.
    N-version programming is argued to create fault-independent implementations~\cite{avizienis85n-version}.
    However, N-version programming does not scale and due to its high additional development costs, it is rarely used in practice.
    In Table~\ref{tab:related-work}, we compare \magic with existing work that tackles the issue of fault-independent implementations.
    Even for Ethereum, which handles billions of dollars in transactions, only a handful of independent implementations exist~\cite{clientdiversify2024, mainnet2024}.
    The implementations are independently developed and maintained; however, despite the risk of billions of dollars, these independent implementations are rarely utilized~\cite{clientdiversify2024}.
    Recent research works try to address this issue for Ethereum by using \acp{TEE} for verifiable client diversity and a reward protocol that incentivizes diverse clients~\cite{javier2025client_diversity}.
    This approach could be integrated into \magic in permissionless setups, but does not address the issue of how to practically get diversified implementations.

    BASE~\cite{rodrigues2001base} provides an abstraction layer that enables the use of independent service implementations, such as different NFS servers.
    Lazarus~\cite{garcia2019lazarus} monitors vulnerability databases, automatically quarantines vulnerable replicas from the system, and patches it with available patches before allowing it back into the system.
    Similarly, FOREVER~\cite{sousa2008forever} uses evolutions of the underlying system (open ports, authentication mechanisms) and updates for the application to diversify replicas.
    Works such as BASE, Lazarus, and FOREVER lighten the burden of diversification from the developer. 
    Garcia et al. have analyzed whether operating systems share vulnerabilities and identified that using diverse operating systems has benefits for intrusion-tolerant systems~\cite{garcia2011os_diversity}.
    Still, these techniques do not provide fault-independent implementations for the \ac{BFT} protocol itself.
    With Proactive Obfuscation~\cite{roeder2010proactive}, the authors propose using obfuscation techniques such as address reordering, stack padding, or system call randomization to diversify replicas.
    Although this approach indeed diversifies the protocol implementation, these techniques can only obfuscate vulnerabilities but cannot remove them; we consider them therefore as an orthogonal approach that can be combined with \magic.
    Recent works, such as SplitBFT~\cite{messadi2022splitbft}, propose a mechanism to ease diversification: the consensus protocol is split into multiple parts (preparation, confirmation, and execution) that can be implemented independently.
    In addition to non-standard hardware (requiring \acp{TEE}), SplitBFT still requires N-version programming. 
    While important, the previously proposed techniques do not meet the goal of automated protocol diversification: they either overlook the \ac{BFT} protocol implementation itself or provide only limited diversification.

\definecolor{goodGreen}{HTML}{a0c15a}
\definecolor{badRed}{HTML}{ff8c5a}
\newcommand{\rwtGood}{\textcolor{goodGreen}{\faIcon{check}}}
\newcommand{\rwtBad}{\textcolor{badRed}{\faIcon{times}}}

\begin{table}[ht]
    \centering
    \small
    \setlength{\tabcolsep}{2pt}
    \renewcommand{\arraystretch}{1.05}
    \begin{tabular}{@{}lccc@{}}
        \toprule
        \textbf{Work}
            & \makecell{\textbf{Diversification}\\\textbf{at protocol level}}
            & \makecell{\textbf{Automated}\\\textbf{generation}}
            & \makecell{\textbf{Code}\\\textbf{diversification}} \\
        \midrule
        SplitBFT~\cite{messadi2022splitbft}              & \rwtGood & \rwtBad  & \rwtGood \\
        Ethereum~\cite{clientdiversify2024,mainnet2024}  & \rwtGood & \rwtBad  & \rwtGood \\
        FOREVER~\cite{sousa2008forever}                  & \rwtBad  & \rwtGood & \rwtBad  \\
        \makecell[l]{Proactive\\obfuscation~\cite{roeder2010proactive}}
                                                         & \rwtGood & \rwtGood & \rwtBad  \\
        Platania et al.~\cite{platanias2014towards}      & \rwtGood & \rwtGood & \rwtBad  \\
        BASE~\cite{rodrigues2001base}                    & \rwtBad  & \rwtGood & \rwtGood \\
        Lazarus~\cite{garcia2019lazarus}                 & \rwtBad  & \rwtGood & \rwtBad  \\
        \midrule
        \magic                                           & \rwtGood & \rwtGood & \rwtGood \\
        \bottomrule
    \end{tabular}
    \caption{Comparison of \magic to existing work. To the best of our knowledge, \magic is the first work that provides diversified implementations for the core protocol of a \ac{BFT} system in a scalable way.}
    \label{tab:related-work}
\end{table}

    \textbf{Rust} is a systems programming language designed to be safe and fast~\cite{matsakis2014rust}.
    It features a strict type system and ownership model to achieve this safety, which has been shown to decrease the number of vulnerabilities~\cite{rebert:2024:google, stoep:2024:google, white-house:2024:report}.
    These safety guarantees have led to the widespread adoption of Rust in distributed protocols and systems programming~\cite{ruesch2019themis, jitaowang2024bftdiagnosis, gkagol2019aleph, bearer2024espresso}.
    But Rust is not infallible.
    With \texttt{unsafe}, blocks of code can be marked that cannot be fully validated by the Rust compiler (e.g., dereferences of arbitrary pointers are allowed).
    This is necessary because the ownership model can be too restrictive for specific use cases.
    However, this can lead to bugs.
    And even without unsafe code, Rust bugs can still occur and lead to security vulnerabilities~\cite{xu2021memory, li2021mir, bae2021rudra, qin2020understanding, hassain2024counterexamples, meneely2025justuserust}.
    Hassnain et al. give concrete examples of such security vulnerabilities in \textit{safe} Rust code~\cite{hassain2024counterexamples} and Meneely et al. show that only up to 58.2\% of C vulnerabilities would have been fixed in a Rust port of the same code~\cite{meneely2025justuserust}.
    Therefore, even for Rust-based agreement protocols, diversification is needed.
    Still, the strong type system and ownership model eliminate many potential bugs and are beneficial for distributed protocols~\cite{ruesch2019themis}.
    For us, the type system and ownership model are also beneficial in identifying that patches are correct, as type conversion errors are caught early by the compiler.

    \textbf{\acfp{LLM}} are statistical models designed to predict text based on previous text~\cite{radford2019language, brown2020language, achiam2023gpt}.
    They have been used for text generation, summarization, translation, and code generation~\cite{brown2020language}.
    This code generation capability is widely used by software developers, with more than 84\% of developers using it~\cite{stackoverflow2025} and 43\% of developers already somewhat trusting the output~\cite{stackoverflow2024}.  

\begin{figure}[]
    \centering
    \includegraphics[width=1\columnwidth]{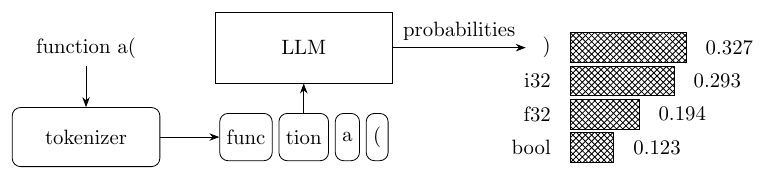}
    \caption{\ac{LLM} code generation example. The \ac{LLM} takes a prompt as input (e.g., \texttt{function a(}) and returns tokens along with their probabilities.}
    \label{fig:llm}
\end{figure}

    An \ac{LLM} is prompted with a set of text that the \ac{LLM} uses to generate the next piece of text.
    Figure~\ref{fig:llm} shows an example of such a prompt (\texttt{function a(}) and the subsequent tokens generated by the \acp{LLM}, along with their probabilities.
    Each token has a probability of being the next token in the sequence, which is controlled by the \textit{temperature} parameter.
    The higher the temperature, the more random the output, i.e., the \ac{LLM} is more likely to generate tokens with lower probabilities~\cite{radford2019language}.
    \acp{LLM} can use \ac{CoT} to reason about tasks before generating outputs~\cite{wei2022chain, kojima2022zeroshot}.
    Recent \acp{LLM} explicitly generate intermediate reasoning steps, sometimes delimited by tags like \texttt{<think></think>}, before a final answer.

     \acp{LLM} have been used for various software engineering tasks, such as understanding code~\cite{nam2024understanding_llm}, generating new code~\cite{mathews2024tdd_llm}, and exploring program design spaces~\cite{zamfirescu2025design}.
    Huynh et al.~\cite{huynh2025vulns} show a 45\% success rate of \acp{LLM} patching vulnerabilities in a dataset of C/C++ vulnerabilities.
    Peng et al.~\cite{peng2025cweval} see similar results (up to 47\%) for vulnerability patching across 5 programming languages.
    The works of Huynh et al. and Peng et al. show \acp{LLM} can fix vulnerabilities in existing code, but do not offer a framework to validate diversification or how to ensure changes are functionally correct.
    Ron et al.~\cite{ron2024galapagos} investigated with Galápagos if \acp{LLM} can be used for automatic N-version programming.
    They implemented automatic validation for the correctness and equivalence of the generated code using off-the-shelf format equivalence checking tools such as alive2~\cite{lopes2021alive2}, or Kani~\cite{vanhattum2022kani}.
    Galápagos verifies the equivalence of the generated code but limits the functions it can diversify to pure functions, which have no side effects.
    This is not suitable for most (if not all) \ac{BFT} implementations that are mostly implemented with stateful functions.
    \magic instead provides an integrated method for \ac{BFT} and \ac{CFT} systems.
    Liu et al.~\cite{liu2023code} and Du et al.~\cite{du2024mercury} both introduced a benchmark for the correctness of \ac{LLM} generated code.
    Liu et al. generate test cases using \acp{LLM} and mutation-based strategies, while Du et al. wrote 1889 Python benchmarks.
    These benchmarks are used to evaluate the correctness of generated code by checking the output of the \ac{LLM} against the fixed expected test case output, but they do not allow verification of any specific function required for generic \ac{BFT} implementations.

\section{Motivation: \acp{LLM} Can Create Patches}
\label{sec:proof-of-concept}

To show the feasibility of our approach, we first conduct a proof-of-concept experiment that evaluates whether \acp{LLM} could remove potential bugs from a reference implementation during diversification.
Although other work has shown this for different systems and languages~\cite{yang2025surveyllmbasedautomatedprogram, huynh2025vulns, peng2025cweval}, we were interested in verifying the capability for two classes of bugs related to our work: faults in distributed algorithms, and security vulnerabilities of Rust.

For the faults in distributed systems we obtained an unpublished bug in Themis\footnote{from private correspondence with the authors}.
For Rust vulnerabilities, we used two real-world security bugs, CVE-2017-1000430~\cite{base64bug} and CVE-2019-16140~\cite{isahcbug}.
This also underscores the fact that security-relevant bugs still pose a risk to distributed protocols, despite the previously mentioned security guarantees for Rust.
For this proof-of-concept we diversified the vulnerable function as described in the design (\Cref{sec:design}) without hints of the vulnerabilities.
We then manually checked whether the vulnerability survived diversification, or whether the \ac{LLM} fixed the underlying vulnerability.

Listing~\ref{listing:decode} details an unpublished impersonation bug in Themis's PBFT implementation.
Messages were not properly checked.
As such, a Byzantine replica could impersonate other replicas.
The fix was to add checks for the source and destination of the message.
In the human fix, an additional optimization is implemented (\texttt{self.decode(src)}) that recursively decodes the rest of the message buffer.
The \ac{LLM} did not detect this because it simply terminates processing on impersonation bugs, which effectively fixes the issue.

The Rust vulnerabilities were also fixed during diversification.
For the first bug, differing from the fix by the library's maintainers, the \ac{LLM} did not change the function signature.
For the second bug, the \ac{LLM} fixed the bug and even removed the need for the \texttt{unsafe} block that introduced the vulnerability, with the drawback of initializing memory twice.

These examples show that \acp{LLM} are able to fix security vulnerabilities in real-world code, including distributed protocols.
This is supported by the works of Huynh et al.~\cite{huynh2025vulns} and Peng et al.~\cite{peng2025cweval}, which show \acp{LLM} successfully patching security vulnerabilities with 45\% and 33\%--47\% probability, respectively.
The fixes demonstrate that \acp{LLM} go beyond the existing automatic code diversification approaches. 
\acp{LLM} are able to fix underlying problems when prompted to diversify code.
Additionally, this proof of concept shows that the changes go beyond simple copy-pasting from the training set of the \acp{LLM}.
For the two Rust fixes, the condition not to change the function signature made the \acp{LLM} generalize beyond the training set.
Furthermore, since the Themis bug is previously unpublished, it is simply not in any \ac{LLM} training set.

Knight and Leveson~\cite{knight1986multiversion} showed in their seminal work that humans make correlated mistakes when creating multiple versions of a program.
While it is unknown which datasets are used for \acp{LLM}~\cite{wan20252025foundationmodeltransparency}, different \ac{LLM} providers use different training and reinforcement techniques~\cite{wang2026reinforcementlearningllmposttraining}, and there is research on interpretability models which can disable parts of the training set during inference~\cite{guidelabs2026steerling} and we have anecdotally seen in this experiment that \acp{LLM} can generalize beyond their training data; still, it is an open question whether different \acp{LLM} can create fully fault-independent code.
Because of this, we focus on ensuring that diverse implementations are representationally different and the compiled binary differs from the original implementation.


\begin{listing}[t]
    \begin{humanbox}
        \begin{minted}[fontsize=\footnotesize, autogobble, linenos=false]{rust}
 if message.destination != self.own_id {
     return self.decode(src);
 }
 if message.source != self.peer_id {
     return self.decode(src);
 }
        \end{minted}
    \end{humanbox}

    \begin{llmbox}
        \begin{minted}[fontsize=\footnotesize, autogobble, linenos=false]{rust}
 if message.destination != self.own_id {
     return Err(io::Error::new(InvalidData));
 }
 if message.source != self.peer_id {
     return Err(io::Error::new(InvalidData));
 }
        \end{minted}
    \end{llmbox}
   \caption{Fix for a decoder bug in Themis where messages were not properly validated.}
    \label{listing:decode}
\end{listing}

\section{\magic}
\label{sec:design}

In short, \magic takes an implementation of a \ac{BFT} (or \ac{CFT}) protocol and automatically generates diversified variants of it (Section~\ref{sec:codegen}).
We do not naively trust the generated code (Section~\ref{sec:system_model}). 
Instead, we validate the generated code using various safeguards.
These validation steps ensure that the generated code is diversified and integrates without manual work into the existing codebase which we verify with extensive stress testing (Section~\ref{sec:safeguards}).
Finally, we show how \magic fits into the lifecycle of distributed systems in Section~\ref{sec:lifecycle}.
In essence, we generate sets of patches, validate them for functional equivalence and ensure they are diverse through safeguards, before using the set of validated patches for the diversification of the system.

\begin{figure}[t]
    \centering
    \includegraphics[width=\columnwidth]{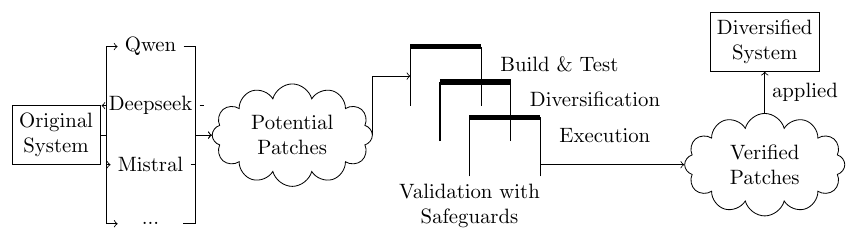}
    \caption{\magic system overview: \magic generates sets of patches which are validated before using the set of validated patches for diversification.}
    \label{fig:barriers}
\end{figure}


\subsection{System and Attacker Model}
\label{sec:system_model}

\begin{figure}
    \centering
    \begin{tikzpicture}

\begin{axis}[
    at={(0.0cm,0cm)},
    width=4.5cm,
    height=4.5cm,
    xlabel={Fault Independence},
    ylabel={Bug Rate},
    ylabel style={yshift=-3mm},
    title={1.3\% \footnotesize Exploit Chance},
    colormap={greenred}{
        rgb255(0cm)=(0,150,0)
        rgb255(2cm)=(255,255,0)
        rgb255(4cm)=(255,0,0)
    },
    point meta min=0,
    point meta max=1,
    enlarge x limits=false,
    enlarge y limits=false,
    xtick={0,0.2,...,1},
    xmin=0, xmax=1,
    ymin=0, ymax=0.15,
    ytick={0,0.05,...,0.15},
    xticklabels={0\%,20\%,40\%,60\%,80\%,100\%},
    xticklabel style={rotate=45, anchor=east},
    yticklabels={0\%,5\%,10\%,15\%},
]
    \addplot[
        matrix plot*,
        mesh/cols=101,
        mesh/ordering=y varies,
        point meta=explicit,
    ] table [
        x=fi, y=br, meta=risk,
    ] {attacker-model-plots/replica_risk_1.3.dat};
\end{axis}

\begin{axis}[
    at={(3.5cm,0cm)},
    width=4.5cm,
    height=4.5cm,
    xlabel={Fault Independence},
    title={3.7\% \footnotesize Exploit Chance},
    colormap={greenred}{
        rgb255(0cm)=(0,150,0)
        rgb255(2cm)=(255,255,0)
        rgb255(4cm)=(255,0,0)
    },
    colorbar,
    colorbar style={
        width=0.2cm,
        title={Risk},
    },
    point meta min=0,
    point meta max=1,
    enlarge x limits=false,
    enlarge y limits=false,
    xtick={0,0.2,...,1},
    xmin=0, xmax=1,
    ymin=0, ymax=0.15,
    ytick={0,0.05,...,0.15},
    xticklabels={0\%,20\%,40\%,60\%,80\%,100\%},
    xticklabel style={rotate=45, anchor=east},
    yticklabels={},
]
    \addplot[
        matrix plot*,
        mesh/cols=101,
        mesh/ordering=y varies,
        point meta=explicit,
    ] table [
        x=fi, y=br, meta=risk,
    ] {attacker-model-plots/replica_risk_3.7.dat};
\end{axis}

\end{tikzpicture}
    \caption{Survival of a whole deployment vs.\ the bug introduction rate \& fault independence against a limited attacker. Even at a 10\% chance of introducing a bug, the system can remain safe even against a strong attacker.}
    \label{fig:expoitchance}
\end{figure}
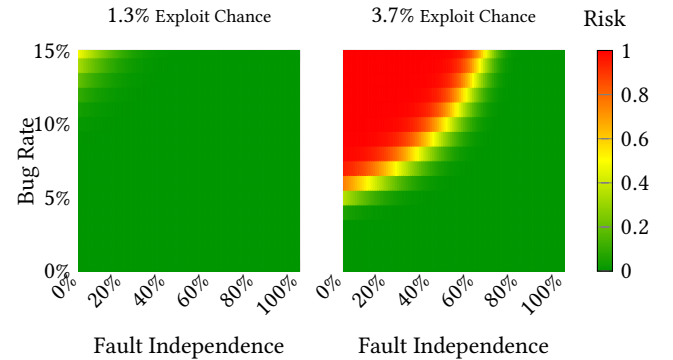

\para{system/attacker model}{
    In reality, a bug in a system does not automatically make it vulnerable to an attacker~\cite{sabottke2015disclosure, jacobs2021epss}. 
    Therefore, in this work we assume that the attacker is not omnipotent in exploiting vulnerabilities, but rather has a probability of exploiting vulnerabilities. 
    This assumption aligns with the works of Sousa et al.~\cite{sousa2006proactive, sousa2006revisited}, who assume a minimum inter-failure time and that the attacker cannot compromise nodes instantaneously or at an arbitrarily high rate.
    
    Our system is designed to diversify a \ac{BFT} protocol implementation. 
	For most current \ac{BFT} deployments, a single implementation is used for all replicas.
	This implementation may contain bugs that an attacker could exploit to compromise more than $f$ replicas.
    To mitigate this risk, the goal of \magic is to generate diversified variants of the \ac{BFT} implementation using \acp{LLM}.
    Code generated by an \ac{LLM} is not always correct in the sense that it exhibits the same functional behavior.
	It may also have other issues and, therefore, needs to be validated before it can be used.
	We assume that the original protocol implementation includes a series of tests to validate it during development.
	These tests are considered the ground truth and correct, but they are not comprehensive; they cannot be used as an oracle to determine if any code is correct or not.
	We use them to validate the generated code and aim to identify and reject all instances where an \ac{LLM} generated incorrect code.
    In rare exceptional cases in which safeguards fail to detect issues, the BFT protocol can tolerate bugs specific to individual replica instances rather than affecting all replicas in the absence of diversification, as shown in~\Cref{fig:expoitchance}.
	Furthermore, approaches such as proactive recovery~\cite{castro03base, castro2002proactive} can be used to strengthen resilience further.
}

\textbf{Implications of the attacker model}
As a result of our attacker model, even if diversification potentially adds (dependent) faults to diversified functions, it does not automatically allow an attacker to exploit the whole system. 
This perspective of considering the safety of the entire system and not just single functions is shown in \Cref{fig:expoitchance}. 
The figure gives the probability that the whole system, not a single replica, is vulnerable, depending on the rate at which faults are introduced during diversification, the probability that bugs are shared during diversification, and the exploit chance, i.e., the probability that an attacker will exploit any given bug.
For the probability of an attacker exploiting a bug, we used numbers from the literature which give probabilities of 1.3\%~\cite{jacobs2021epss} and 3.7\%~\cite{sabottke2015disclosure}.
Each replica has 200 functions in which a vulnerability would be severe enough to overtake the replica.

The analysis shows that even in a scenario in which 3.7\% of bugs are found and exploitable, the diversification will still protect the system even if up to 10\% of the functions introduce faults and at least 90\% fault independence exists between them. 
This gives us the assurance that as long as we can keep the exploit rate low enough, potential bugs in diversification will not risk the whole system.
Note that while our work cannot show fault-independence, we show representational (using clone detection) and binary diversification, since even with full fault correlation between replicas the system can remain safe as seen in the lower left quadrants of \Cref{fig:expoitchance}. 
With recent advancements in \acp{LLM}, \acp{LLM} have been used to identify long standing vulnerabilities in projects such as Firefox, OpenBSD, or ffmpeg~\cite{anthropic2026mythos_red, anthropic2026mozilla}.
We imagine that this capability can identify bugs in implementations and patches and thus reduce exploit probabilities for whole systems when used before deployment of systems.
Further, recent scientific works use \acp{LLM} to identify and fix security vulnerabilities with fuzzing~\cite{zhang2026ai_fuzz} which could further reduce vulnerabilities from implementations.

\subsection{Code Generation}
\label{sec:codegen}
To motivate the code generation design, we first answer questions about how to use \ac{LLM} for diversification.

\textbf{At what level should we diversify?}
There are multiple levels at which we could diversify the code: a single line at a time, at the function level, at the file level, or at the project level.
We decided to diversify at the function level.
Changing one line at a time is too fine-grained and would leave little room for meaningful diversification.
On the other extreme, changing the entire project at once would be challenging for \acp{LLM} as it reaches the limits of how much code can be generated.
We decided against diversifying at the file level and instead diversifying at the function level.
Otherwise, a change might affect multiple functions at once, creating functional dependencies between patches.
This would mean that for the application of patches we could no longer freely mix and match patches for different functions from different \acp{LLM}.
For sufficient context on the codebase, we provide the \ac{LLM} with additional information about the function (e.g., function signatures and struct definitions).

\textbf{Should we explicitly prompt the \ac{LLM} with the baseline code or just the intended functionality?}
This way, we would not risk bugs from the baseline code being copied.
Unfortunately, we found that this approach does not work for two reasons.
Firstly, the original code lacks sufficient comments and explanations to explain its intended functionality clearly (under-specification problem).
Secondly, even when manually providing more context for this approach, the \ac{LLM} would generate code that would not fit into the existing codebase.

\textbf{How to create multiple diversified versions of the code?}
We use multiple \acp{LLM} to generate diversified code.
This way we use \acp{LLM} implemented with different algorithms and trained with different training data.
While \acp{LLM} can share public training data, e.g., data from GitHub, each \ac{LLM} is trained with a different focus, e.g., explainability or usability.
\acp{LLM} are fine-tuned to fit this focus, which is often a proprietary process unique to each \ac{LLM}.
In doing this, we diversify the risk of a single \ac{LLM} potentially generating code with the same bug.
We not only generate diversified code, but we also use diversified \acp{LLM} to generate the code.
For each newly generated variant of the code, we validate that it is diversified compared to the original code and the previously generated versions.
More specifically, we check if the code is not a clone~\cite{roy2007survey} and if the generated binary differs from any previous code.


\begin{figure}[t]
\begin{tcolorbox}[
  enhanced,
  promptbox,
  title={Diversification Prompt},
  label={lst:pbft-prompt},
  overlay={
    \node[
      anchor=south east,
      xshift=-1.2mm,
      yshift=1.2mm,
      inner sep=0pt
    ] at (frame.south east) {
      \begin{tikzpicture}[x=1mm,y=1mm]
        \draw[
          rounded corners=0.7mm,
          draw=black!65,
          fill=white,
          line width=0.35pt
        ] (0,0) rectangle (9.2,5.8);

        \draw[draw=black!55,line width=0.3pt]
          (0,4.15) -- (9.2,4.15);

        \node[
          anchor=west,
          font=\ttfamily\tiny,
          inner sep=0pt,
          text=black!80
        ] at (1.0,2.25) {\$};

        \draw[
          draw=black!80,
          line width=0.45pt,
          line cap=round
        ] (3.0,1.55) -- (5.9,1.55);
      \end{tikzpicture}
    };
  }
]
\scriptsize\ttfamily
> You are a senior Rust engineer that will create a diversified function implementation. You will be given a function from a practical byzantine fault tolerant (PBFT) implementation. Consider if the function is implemented correctly based on the function name and the PBFT specification. Write correct code, if there is a bug in the code provided you will fix the bug in the output. Your goal is to write correct and safe code. Avoid unsafe code if possible, do not use unwrap. Write defensive code. You can add more checks and assertions if you think they would be useful. The code you generate will have: * the same function signature * return the same type The code will have to work as a drop in replacement, only the internals of the function can change. The function you diversify will be provided in <function> </function> tags. You will output the diversified function in <output></output> tags. The output should only contain valid code. Do not add ```rust in the output. You only return the newly created rust code, that can compile without warnings. No other text, explanations or other information. Just Rust Code! Here is the function <function> \{FUNCTION\} </function>. For context here are all the functions the function calls <called> \{CONTEXT\_FUNCTION\} </called>. For context here are all the data structures the function uses <structures> \{CONTEXT\_DATA\} </structures>. Task: Provide an alternative and safe Rust implementation with the same function signature and functionality as the given code \\ snippet in <function> tags and output it with <output> tags!
\end{tcolorbox}
\caption{General PBFT prompt for the \ac{LLM}. For other protocols, protocol-specific terms are replaced accordingly.}
\label{listing:prompt}
\end{figure}

\textbf{Prompts.} 
We prompt the \ac{LLM} with the full-function body that we want to diversify.
Our general prompt is shown in Listing~\ref{listing:prompt}.
Additionally, we provide the function signature and struct definitions used in the function.
This additional context allows the \ac{LLM} to generate more meaningful code.
An example excerpt of the context is shown in Listing~\ref{listing:context}.
With this context, the \ac{LLM} can access struct definitions not previously seen in the original function.
The evaluation (Section~\ref{sec:evaluation}) demonstrates that this approach produces more successful patches.
The prompt limits what the \ac{LLM} will generate.
It ensures that only code is returned, with no explanations or comments on the code.
This way, we can ensure that the generated code integrates into the existing codebase without requiring any post-processing.
Without these limitations in the prompt, we found that the \ac{LLM} would often deviate from the task generating output other than code.

\begin{listing}[]

    \begin{codebox}[Dynamically generated LLM context]
        \begin{minted}[fontsize=\footnotesize, autogobble, linenos=false]{rust}
enum ViewState {
    Regular,
    ViewChange {
        new_view: u64,
        _timer: Timer,
...
pub struct OrderingLog {
    current_view: Slots<OrderInstance>,
    old_views: Vec<Slots<OrderInstance>>,
}
        \end{minted}
    \end{codebox}

    \caption{The context for the LLM is dynamically generated from the diversified function. We provide definitions of any structs or enums used in the function, as well as function signatures.}
    \label{listing:context}
\end{listing}

\subsection{Code Safeguards}
\label{sec:safeguards}

Before using any generated code, we need to validate that the code behaves as expected and is diversified.
Therefore, the generated code must pass through multiple safeguards before being used (see Figure~\ref{fig:barriers}).
In particular, we validate that the code is functionally correct as well as an actual diversification of the code.
For the functional tests we build the code to check syntactical correctness, run unit tests, and run system tests together with unpatched replicas, before finally stress testing the patches by running fully patched implementations together.
Further, to validate diversification we use clone detection on the level of the source code as well as validate that the generated binary code for the function differs from the unpatched implementation.

\textbf{Build \& Test.}
The initial validation is the build-and-test safeguard.
This safeguard is as simple as it sounds; we try to build the code and run all tests.
After this safeguard, we know the code is syntactically correct, and all tests continue to pass.
As with conventional development, high test coverage helps catch obvious errors early on. 

\textbf{Diversification.}
After verifying that the code is syntactically correct, we validate that it is diversified.
For this we use clone detection~\cite{roy2007survey, zhu2022msccd} and compare the generated binary code for the function.
Clone detection is widely researched with four types of clones considered:
\begin{description}
    \item [Type 1:] Identical code except for whitespace and comments
    \item [Type 2:] Syntactically identical code except for changes in variable/function names, types, layout and comments
    \item [Type 3:] Statements can be changed, added or removed in addition to Type 1 and Type 2 changes
    \item [Type 4:] Code that performs the same computation but is implemented through different syntactic variants
\end{description}
The goal of \magic is to achieve type 4 clones, that is, code that implements the same functionality but through different syntactic means.
For example, consider a reference implementation of the Fibonacci function.
A type 3 clone might switch around additions, add helper variables, or change types of variables, which is obviously not a significant change and is something we want to avoid.
A type 4 clone, on the other hand, might re-implement the function from an iterative to a recursive form.
But ensuring that a patch is not a clone is not enough.
Compiler optimizations can generate the same machine code even for some type 4 different code.
E.g., Rust's zero-cost abstractions trade compile-time effort for turning higher-level language features into efficient machine code~\cite{rust_zero_cost_abstractions}.
Therefore, we also compare the machine code of the original code with the generated assembly of the diversified code.
For this, we compile the code once unpatched and with the generated patch.
We then compare the generated binary for the function we diversified.
We accept the patch only if the function's assembly differs between the two binaries.

\textbf{Execution Check.}
The execution check is divided into two phases.
First, we introduce a fast and simple check where a single patch is verified alone with unmodified replicas; second, we validate multiple patches together to identify faults that only occur if multiple patches interact with each other.
This is inspired by software engineering practices where stabilization branches are used to test multiple patches together for a release~\cite{castelluccio2019empirical} and work that shows that testing multiple patches together achieves more cost effective bug detection~\cite{najafi2019bisecting}.

In the first execution check, we execute a single replica with the diversified function together with multiple unchanged replicas for a fixed period.
We set this period to be long enough for requests to be committed, checkpoints to be created, and (after a deliberate crash of the leader) a view change to happen.
Here, we verify that the replica behaves as expected, e.g., it does not crash and participates in the consensus process. 

In the second phase, the \textbf{stress test} phase, once we have enough patches to get diverse fully patched replicas we test the fully diversified replicas.
Again, we ensure that a view change occurs by deliberately crashing the leader.
If we identify faults in this step, they could be caused by a single failed patch or by multiple patches that fail together.
Patches that fail on their own should have been identified before, but, as the runtime is parameterizable, they might only be identified with the additional time spent testing in this test with four fully diversified replicas.
For these single failing patches, we narrow down the potential set of patches by repeatedly bisecting the set into two halves, which we test independently until one patch remains.
In the second case, where multiple patches only fail in conjunction, we can not identify them with the bisecting routine as we can't ensure that the set of responsible patches remains in each bisection.
Instead, we first identify all the patch-replica pairs that are present in every faulty round, e.g., if we repeat the stress test 100 times and 4 rounds fail, we then find all the patch-replica pairs present in all 4 faulty executions.
We use patch-replica pairs, as it matters which patch was on which replica, e.g., it matters if the patch was on the leader replica rather than on the follower replica.
Next, we create k-subsets for $k=2,..,10$ from the set of the potential patch-replica sets identified and check for all k-subsets whether any specific k-subset occurs in a round without fault.
We repeat this with increasing $k$ until we find sets of patches which were never part of a round without fault.
We execute these subsets again, discarding them from a final deployment if a fault occurs during execution again.

After passing these safeguards, we have ensured that the code has been successfully diversified, and we have high confidence that it will behave as expected.
We can now use the diversified function.

\subsection{How to use \magic \& Lifecycle Management}
\label{sec:lifecycle}

Software is never fully finished, as new requirements and bugs are constantly found and addressed.
We imagine that before the initial deployment of a distributed system \magic is used to create diversified implementations of the system resulting in some set of patches.
There are two scenarios for changes that affect our system, (1) function local changes and (2) more global changes.
For (1) local changes, users of \magic could identify the functions that have changed and discard all previously created patches for that function, making new ones while retaining the patches for all other functions.
For (2) global changes, e.g., to some widely used data structures, we recommend discarding all previously created patches.
\magic decreases the cost of diversification by orders of magnitude compared to manually creating diversified implementations (see Section~\ref{sec:evaluation}), and thus discarding all previously created patches before recreating new ones is feasible.

Proactive recovery of replicas can prevent the accumulation of faults by proactively replacing replicas with new versions~\cite{castro2002proactive}.
Using the cost efficiency of \magic, it can also be used to continuously diversify a system at runtime.
Replicas can be shut down and then restarted with new patches applied.

\section{Implementation}
\label{sec:implementation}

We have implemented our approach for the Rust-based \ac{BFT} implementations PBFT Themis~\cite{githubthemis}, HotStuff~\cite{githubhotstuff} and the \ac{CFT} Raft implementation Openraft~\cite{databend2024openraft}.
\proj is not limited to these protocol implementations, and it can be applied to any Rust project which has unit tests (optional) and a system test that verifies that the system is working as expected (e.g., is making progress).
As an anecdote, once implemented for Themis, we were able to integrate Openraft and HotStuff within $\approx$5 hours of work.
\magic makes no special requirements on the \ac{LLM} for the diversification.
For inference we used infrastructure provided by the GWDG~\cite{ali2024chatai}.
We extracted functions and added context to functions using the parsing tool TreeSitter~\cite{githubtreesitter}.
Inference on the \acp{LLM} was done sequentially, but could also be parallelized.

\textbf{Themis \& Openraft \& HotStuff.}
We have implemented and evaluated \magic for three different distributed algorithms: PBFT, Raft, and HotStuff.
With this we evaluate \magic for PBFT as a classical \ac{BFT} protocol, HotStuff as a new blockchain protocol and Raft as a \ac{CFT} protocol, for each limiting the diversification to the protocol itself.
For more comprehensive diversification, all the code could be diversified.

\textbf{Safeguards.}
We implemented the aforementioned safeguards to validate patches.
Each safeguard was executed and the results were stored, i.e., even if the ``test'' safeguard failed, it was still checked if the patch is a clone or not and whether or not the binary is different.
The \textbf{build \& test} safeguards are simple \texttt{cargo build} and \texttt{cargo test} commands.

For \textbf{clone detection} we used the work of Zhu et al. called MSCCD~\cite{zhu2022msccd}.
With MSCCD we detect type 3 clones.
For this we used the default configuration of MSCCD with a detection threshold of 0.7.
This can be tuned where a higher threshold increases accuracy, but reduces recall.
It should be noted that this is a conservative safeguard, e.g., if 8 lines out of 55 were copied, then the whole function is marked as a clone.
Additionally, it should be noted that MSCCD can have both false positives as well as false negatives~\cite{zhu2022msccd}, potentially identifying non-clones as clones.

After the clone comparison, we use Bean~\cite{bean}, a binary analysis tool for the \ac{ELF}~\cite{elf}, to \textbf{compare the binaries}.
Here, we compare the generated code for the diversified functions, examining the original and diversified binaries.
A single-bit change does not necessarily result in a change in the binary as measured by Bean. 
Bean skips all NOP instructions, alternative encodings of the same operand (8-bit versus 32-bit displacement) result in the same hash. 
The hash is position-independent, with branch and RIP-relative targets resolved relative to the function start. 
The (mangled) name, binding, and section are excluded from the hash. 
This makes Bean robust to single-bit changes.
If a hash over the function's binary code is different, we consider the diversification successful.

For the \textbf{execution safeguard}, we execute the patched code with three unpatched replicas.
We expect the overall system to make progress (even if we deliberately crash one of the replicas), i.e., the system should process client requests, and replicas should not crash.
If this is the case, we consider the patch successful.
For this we execute the replicas and client on one machine.
We use one client configured to oversaturate the replicas.

Finally, for the \textbf{stress test safeguard}, we used Shadow~\cite{rob2022shadow}, a discrete-event network simulator with a virtual clock and a deterministic, seeded scheduler, and executed four replicas, each fully diversified from the set of patches that passed all previous safeguards.
We configure Shadow to inject faults after a configurable time; its deterministic execution and reproducible fault injection are what let us identify problematic patches, as we detail in \Cref{sec:evaluation}.

\textbf{Diversification of the Implementation.}
After all safeguards including the stress test, we have a set of patches that are functionally equivalent but are type 4 clones.
These patches are then applied to the codebases for diversified replicas.
For each function that was diversified, we take a random patch from any of the \acp{LLM} and apply it to the codebase.
This way, we ensure that the diversified replicas are distinct from one another and that we do not rely on any particular \ac{LLM}.

\section{Evaluation}
\label{sec:evaluation}

\newcommand{\qpercent}{Q\textsubscript{loc}}

\newcommand{\qsafeguard}{Q\textsubscript{s.g.}}

\newcommand{\qeffect}{Q\textsubscript{eff}}

\newcommand{\qmul}{Q\textsubscript{unq}}

\newcommand{\qprompt}{Q\textsubscript{ctx}}

\newcommand{\qstress}{Q\textsubscript{stt}}

\para{Intro}{
    In this section, we evaluate the effectiveness of our approach. We answer the following questions: 
    \begin{itemize}
        \item[\qeffect] How effective is our approach in diversifying code?
        \item[\qsafeguard] At what safeguards do patches fail?
        \item[\qmul] How many new (unique) patches are created if we request multiple variants of the same \ac{LLM}?
        \item[\qprompt] What is the effect of specific prompts on the results?
        \item[\qpercent] What percentage of a codebase can we diversify? 
        \item[\qstress] How do individually verified patches behave in a (long-term) stress test with fully diversified implementations?
    \end{itemize}
}

\textbf{System Configuration.}
\para{Setup}{
    For the diversification we use the open models of Mistral~\cite{mistral_model}, Qwen3~\cite{qwen3_model}, DeepSeek-R1~\cite{deepseek_model}, as well as the commercial model of Sonnet~\cite{claude}.
    We ran the experiment on CloudLab~\cite{dimitry2019cloudlab} on the \texttt{d6515} machine type unless otherwise stated.
    The machines have 32 cores (AMD 7452 at 2.35GHz), 128GB of RAM (8x 16 GB 3200MT/s RDIMMs), and 1TB of disk space.
    For inference we used inference servers from the GWDG~\cite{ali2024chatai}.
    We have limited our evaluation to functions with at least 10 \ac{LoC}.
    This was done, since with fewer \ac{LoC}, there is little opportunity for diversification.
}

\begin{figure*}[]
    \centering
    \begin{subfigure}{0.30\textwidth}
        \centering
        \includegraphics[width=\textwidth]{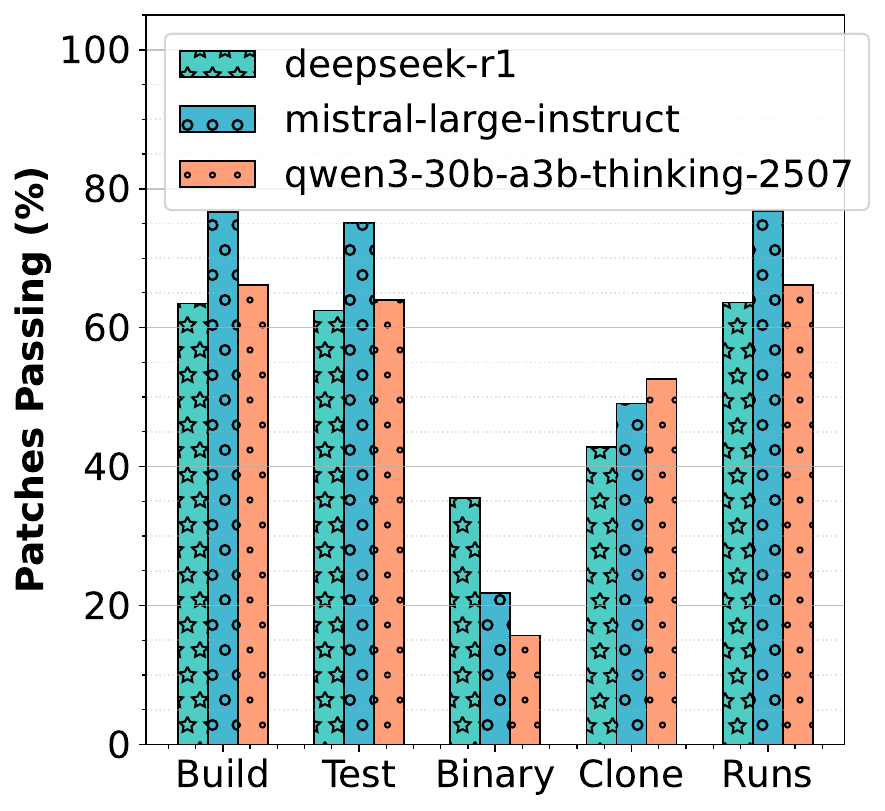}
        \caption{PBFT}
        \label{fig:new-eval-pbft}
    \end{subfigure}
    \hfill
    \begin{subfigure}{0.30\textwidth}
        \centering
        \includegraphics[width=\textwidth]{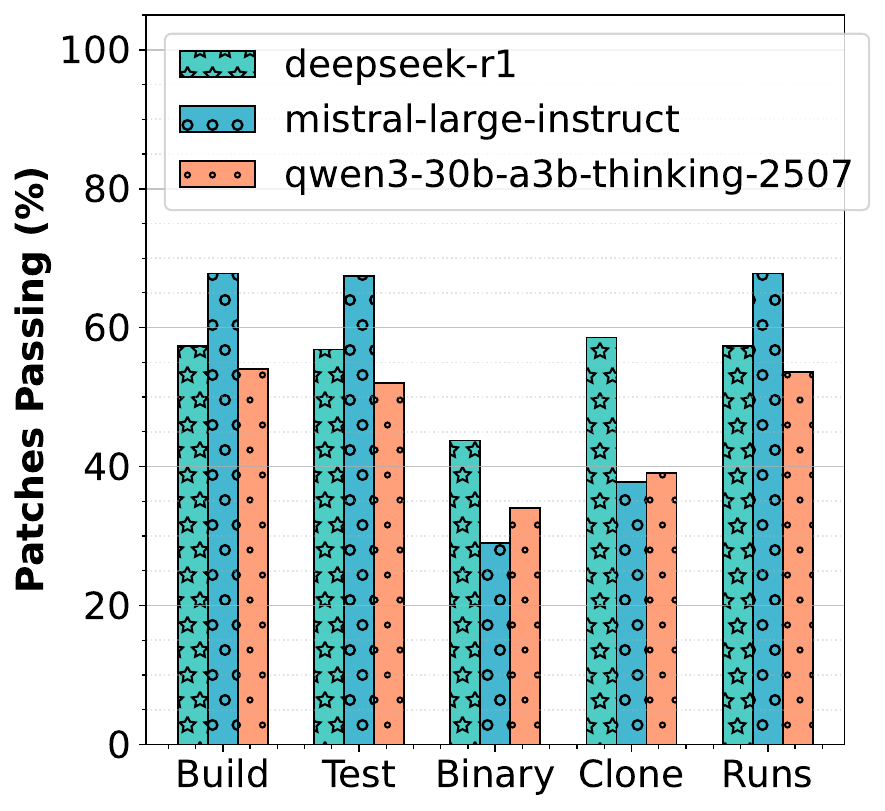}
        \caption{HotStuff}
        \label{fig:new-eval-hotstuff}
    \end{subfigure}
    \hfill
    \begin{subfigure}{0.30\textwidth}
        \centering
        \includegraphics[width=\textwidth]{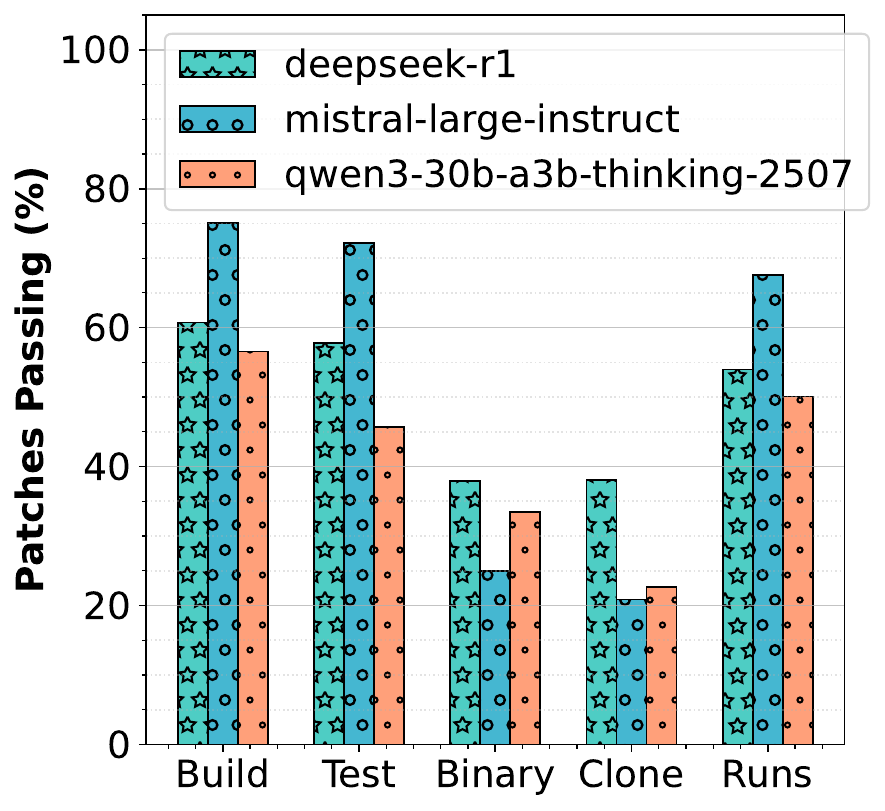}
        \caption{Raft}
        \label{fig:new-eval-raft}
    \end{subfigure}
    \caption{Safeguard results for PBFT, HotStuff, and Raft. Each safeguard was tested, even if previous safeguards failed.}
    \label{fig:safeguards}
\end{figure*}

\begin{figure}[]
    \centering
    \includegraphics[width=0.9\columnwidth]{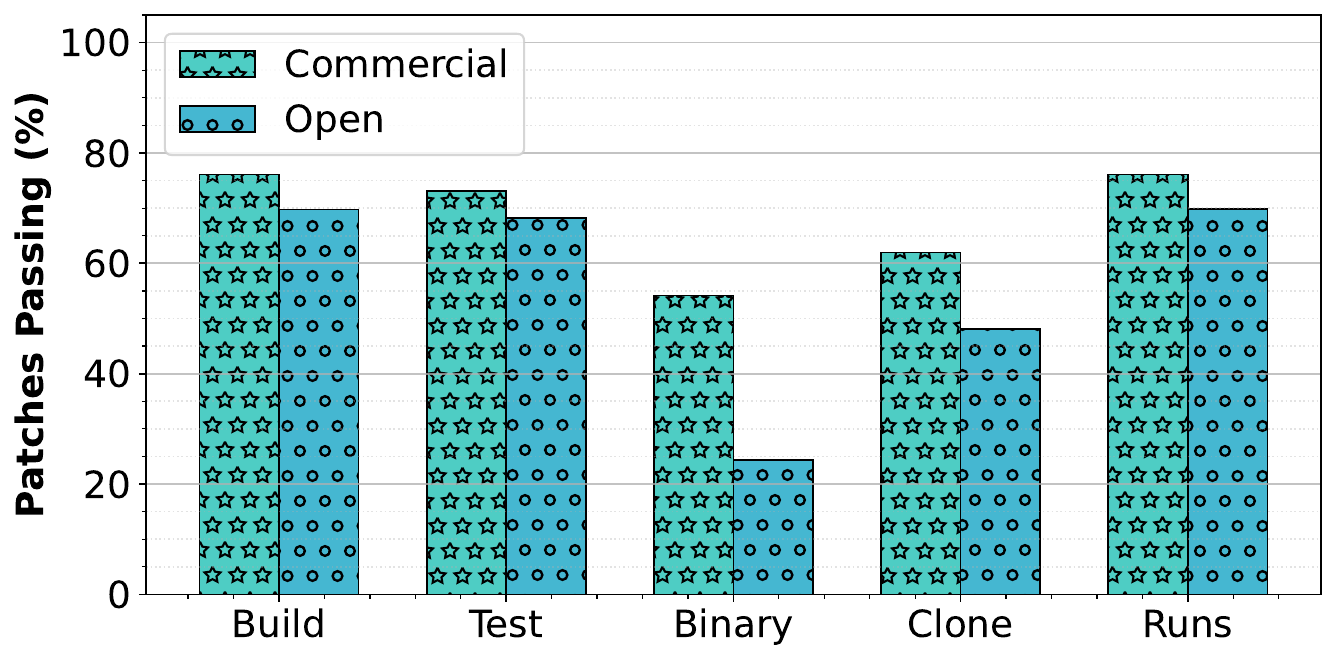}
    \caption{Results for PBFT with the averaged results of the open models Qwen3, DeepSeek, and Mistral compared to the commercial model Sonnet 4.5 from Anthropic.}
    \label{fig:commercial}
\end{figure}

\textbf{Diversification of PBFT, HotStuff, and Raft.}
Across all experiments we have generated 4945 patches for our evaluation.
Not all of these patches are successful patches that can be used in production.
Figure~\ref{fig:safeguards} shows the results of the safeguards on these patches observed.
It should be noted that these probabilities are not independent.
A patch which correctly builds also has a high chance of passing the tests and passing the runtime check ($\frac{1201}{1224} \approx 99.5\%$ for PBFT, $\frac{885}{992} \approx 89.2\%$ for Raft, and $\frac{454}{468} \approx 97.0\%$ for HotStuff).
Similarly, the clone detection and binary difference checks are dependent on each other: if one succeeds, the respective other one also passes with a high probability (not as easy to calculate\footnote{All patches that run can also be built, but patches with a different binary can come from detected clones (subset of the \ac{LoC}), and identical binaries can come from type 4 different patches with compiler optimizations.}).

DeepSeek and Qwen3 automatically generate \acf{CoT} (as outlined in the background \Cref{sec:background}), while Mistral does not.
This affects their code generation: the reasoning models explored more diverse solutions, while the non-reasoning model generated more conservative outputs closer to the input.
Figure~\ref{fig:safeguards} illustrates this trade-off: Mistral produces less diverse code (lower binary/clone safeguard pass rates) but achieves higher correctness (build/test/run pass rates).

For comparison we have also investigated the results of a state of the art commercial model against the combined results of the open models.
As we see in Figure~\ref{fig:commercial} the commercial model achieves better results across the board, especially for the diversification safeguards.
As a fraction, 6.5\% of all patches generated by open weight models pass all safeguards compared to 19.4\% for the commercial model.
For the rest of the evaluation we will use the open-weight models.
This gives us a lower bound on the effectiveness of \magic and also enables the reproducibility of the results.
In production we imagine the use of state-of-the-art models to yield even better results than shown here.

As runtime failures are quite rare (that is, for patches that build, i.e., if a patch builds then it also has a high chance of passing tests and passing the runtime safeguard), we highlight one that caused PBFT to misbehave.
Listing~\ref{listing:crash} shows an excerpt of a modification to the \texttt{is\_duplicate} function that caused the program to misbehave, which was counted as a crash.
In the original code, the function verifies that the request has a sequence number.
This is ignored by the patch, which only checks if the request hash is already in the set of requests.
At runtime, this causes the replicas to timeout after a view change.
This timeout is interpreted as a crash and thus failure of the safeguard.

We calculated the cost by multiplying the cost per input and output token per \ac{LLM} with the average number of tokens used.
The cost of generating patches is extremely low, much cheaper than manually implementing them.
One single successful patch costs less than 10 cents.
As an anecdote, the cost of running all experiments for this paper, including the evaluation and development of the idea with commercial model providers, was less than 50 dollars.

\answer[\qeffect \qsafeguard]{\magic can generate diversified patches}{
    If a patch builds, it has a high chance of passing all safeguards.
    While state-of-the-art models achieve better results, open-weight models offer a reproducible lower bound
}


\begin{listing}[]

    \begin{codebox}
        \begin{minted}[fontsize=\footnotesize, autogobble, linenos=false]{rust}
- self.requests.get(&digest)
-              .and_then(|e| e.sequence)
-              .is_some()
+ let digest = H::hash_request(&request);
+ let is_duplicate_request = self.requests.contains_key(&digest);
        \end{minted}
    \end{codebox}
    \caption{Excerpt of a modification to the \texttt{is\_duplicate} function that caused the program to misbehave.}
     \label{listing:crash}
\end{listing}

\begin{figure}[]
    \centering
    \begin{subfigure}{0.8\columnwidth}
        \centering
        \includegraphics[width=\textwidth]{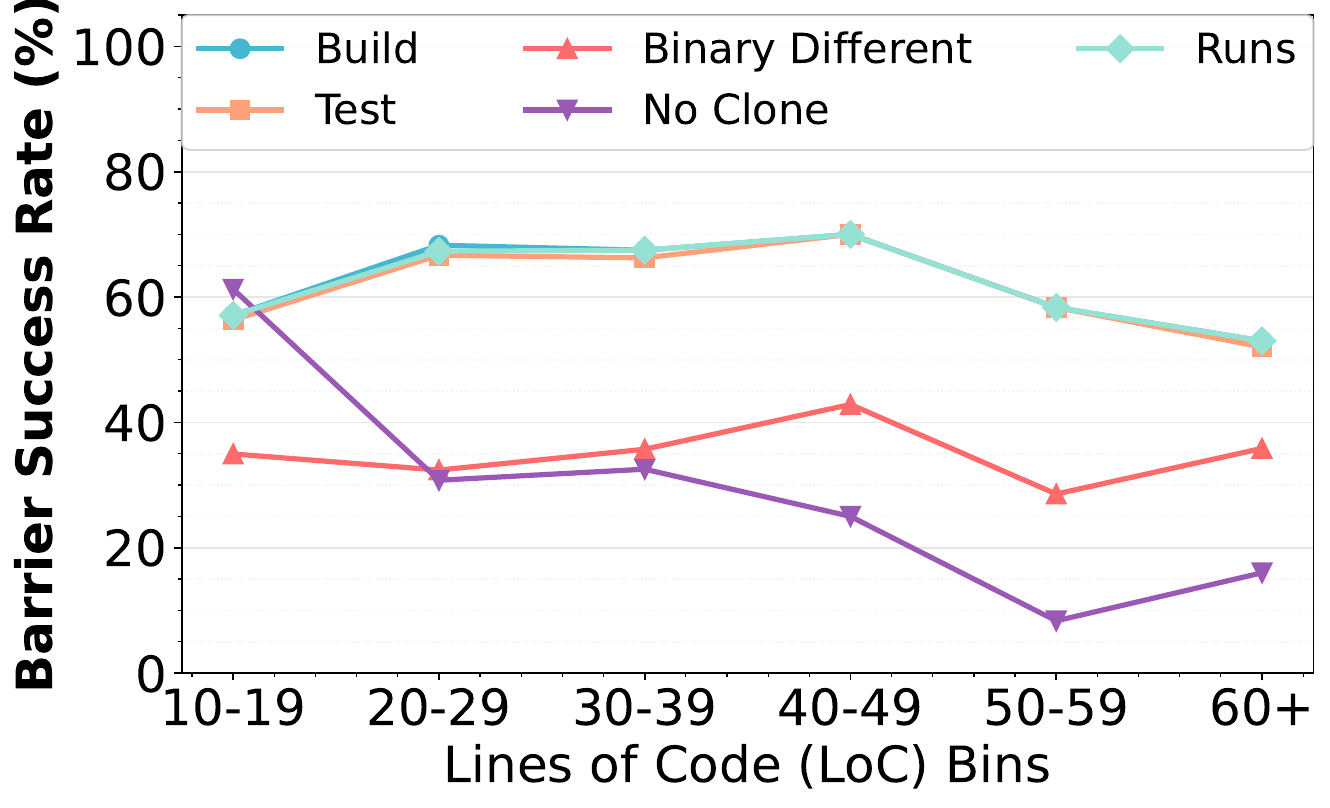}
        \caption{HotStuff: we see the build and runs line overlap as the runtime check is less strict than the test safeguard.}
        \label{fig:loc-barrier-hotstuff}
    \end{subfigure}
    \hfill
    \begin{subfigure}{0.8\columnwidth}
        \centering
        \includegraphics[width=\textwidth]{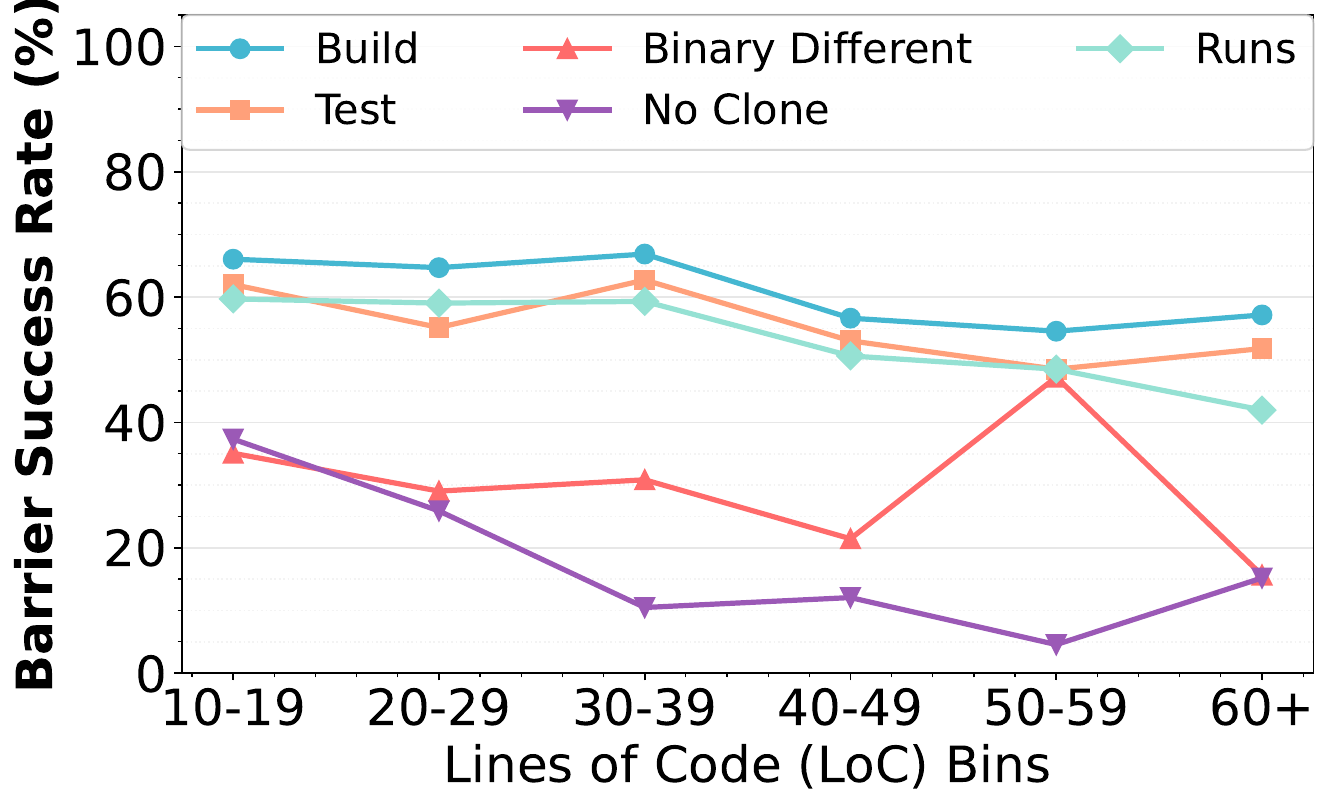}
        \caption{Raft: the uptick for the binary difference in the 50-59 bin is explained by the small sample size with 9/18 succeeding.}
        \label{fig:loc-barrier-raft}
    \end{subfigure}
    \caption{Safeguard results broken down by lines of code bins. Even for big functions with 60+ lines of code we see patches being generated. PBFT is omitted for space reasons, it shows a similar trend.}
    \label{fig:loc_bins}
\end{figure}

In Figure~\ref{fig:loc_bins} we see the success rate of the safeguards broken down by the lines of code per function diversified.
This combines the patches of all \acp{LLM} for each protocol.
In general, we see no difference in the success rate for the safeguards based on the lines of code.
Only for clone detection, we see the success rate decrease with increasing lines of code.
This is explained by the fact that in a large function, there are more chances that parts of the function are a clone.
As explained in \Cref{sec:implementation} we used the default configuration of MSCCD, where even a subset of copied lines marks the function as a clone even if the remaining changes are substantial.
For the binary difference, we do not see this effect, as in the remaining 55 changed lines there might be substantial changes.
Thus, our configuration might be considered the strict baseline, which could be relaxed based on practical considerations.

\answer[\qeffect]{\acp{LLM} can generate patches for large functions}{
    The success rate of patches decreases with complexity, though this is mostly explained by the specific clone detection settings used.
    Still, even for big functions with more than 60 lines of code there are successful patches generated.
}

\begin{figure*}[]
    \centering
    \includegraphics[width=0.83\textwidth]{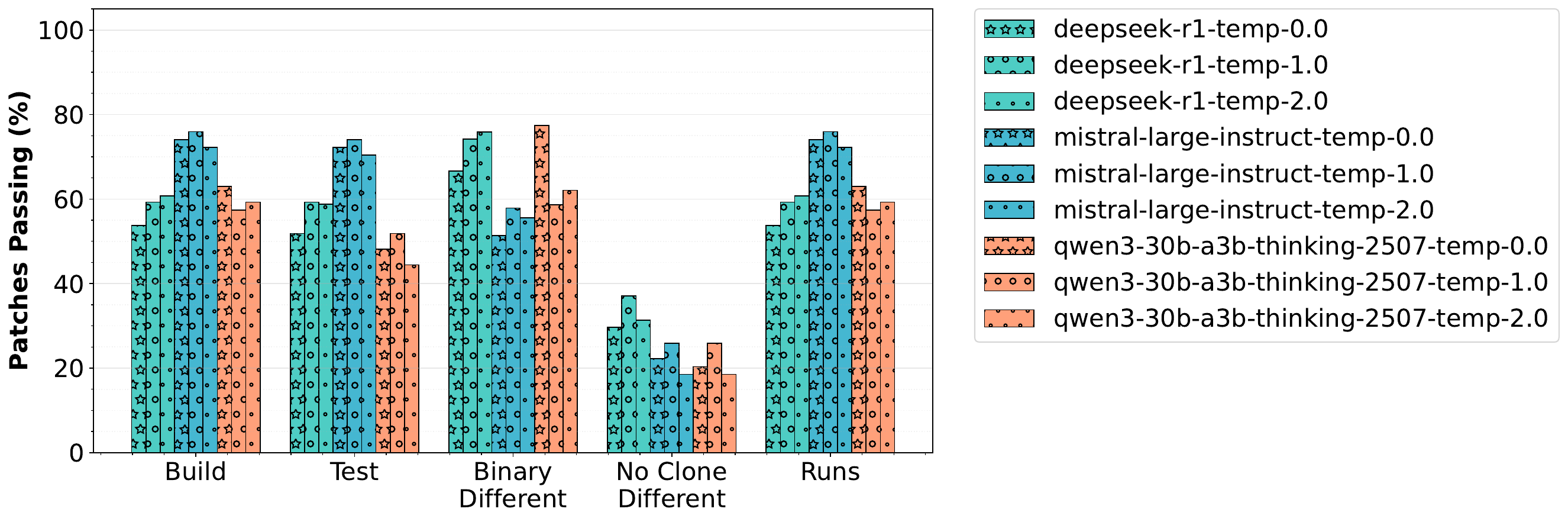}
    \caption{Safeguard results with three different temperature settings for PBFT. Temperature has little effect on the results.}
    \label{fig:llm-temperature}
\end{figure*}

\textbf{Temperature.}
We found that the temperature used for \ac{LLM} has little effect on the number of successful patches generated.
Figure~\ref{fig:llm-temperature} shows the number of successful patches generated for different temperatures.
We generated patches for PBFT with temperatures 0.0, 1.0, and 2.0.
In Figure~\ref{fig:llm-temperature} we do not see significant differences between the temperature and the success rate between the different safeguards.
This is surprising, especially for higher temperatures, i.e., more ``creative'' \acp{LLM}.
We would have expected the clones to decrease and the binary difference to increase with increasing temperature.
More research is needed to explain this effect.

\answer[\qeffect \qsafeguard]{Temperature: little influence on success rate}{
    Temperature only has a slight effect in either direction on the success rate of safeguards.
}

%


\begin{figure}[]
    \centering
    \includegraphics[width=\columnwidth]{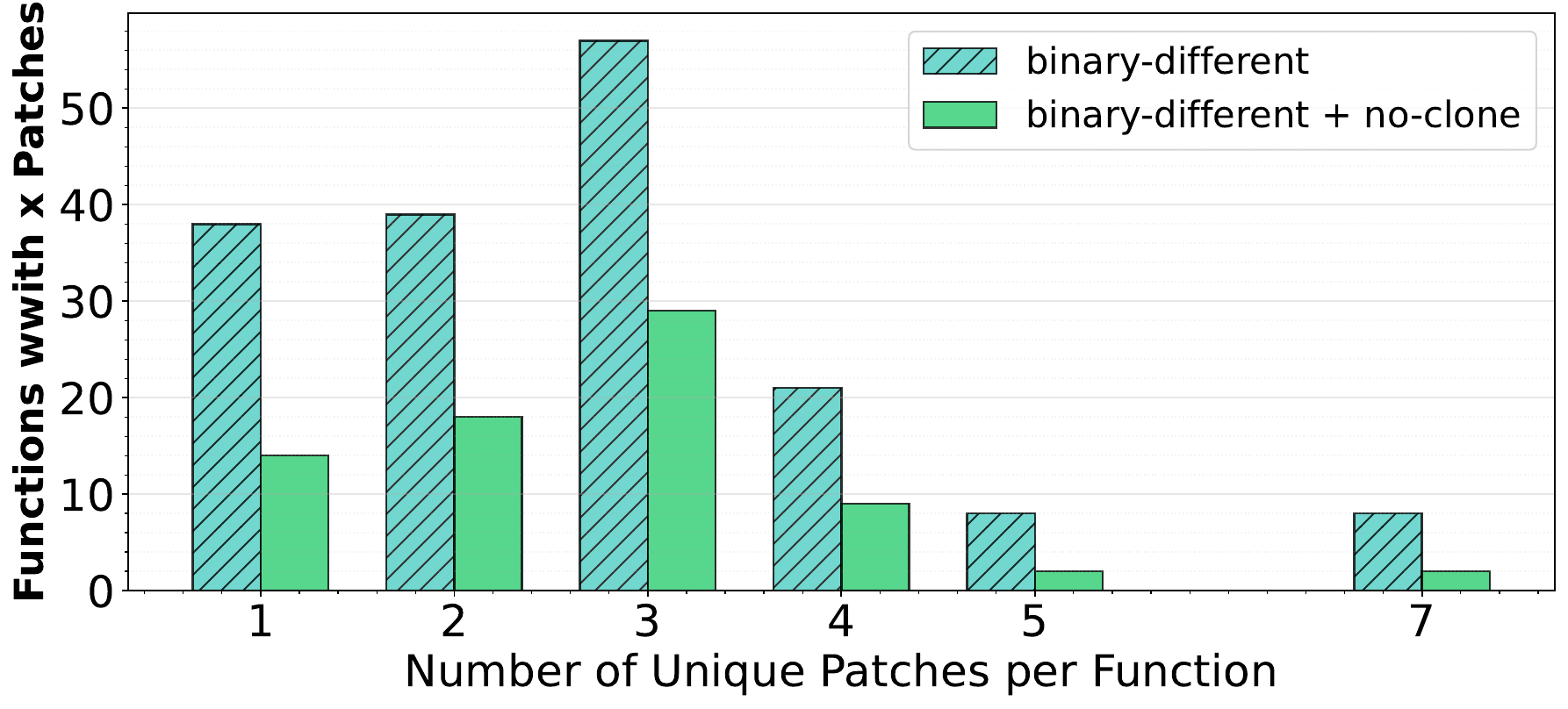}
    \caption{Running \magic multiple times for PBFT results in several patches with all-to-all unique binary + no clone.}
    \label{fig:n-version}
\end{figure}

\begin{figure}[]
    \centering
    \begin{subfigure}{0.9\columnwidth}
        \centering
        \includegraphics[width=\textwidth]{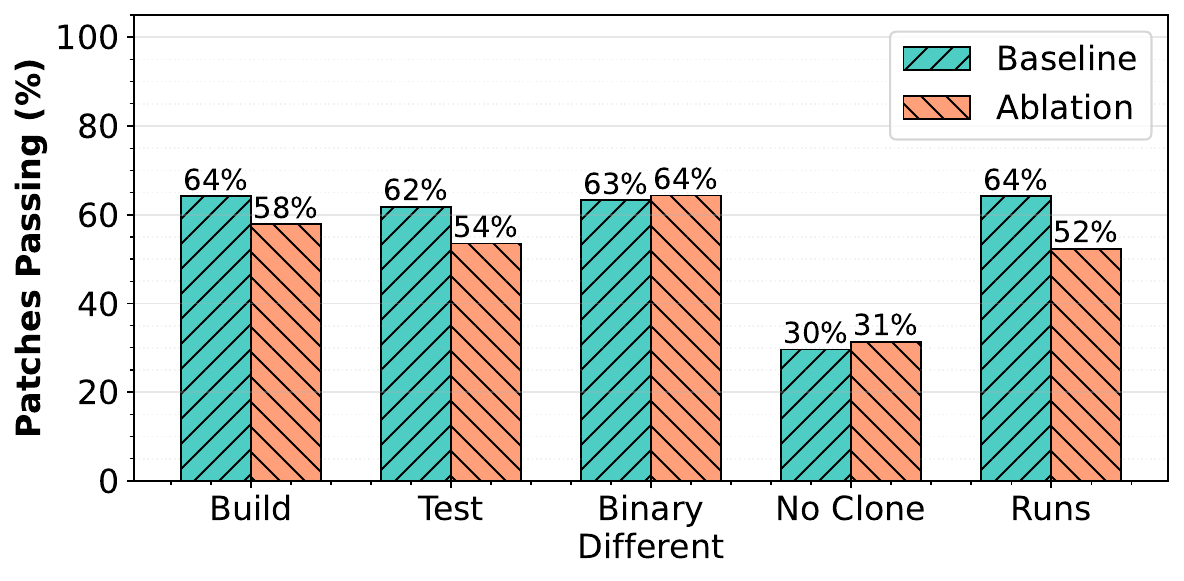}
        \caption{Without context about data structures and functions within the codebase the \acp{LLM} hallucinate about non existing elements in the codebase.}
        \label{fig:ablation_context}
    \end{subfigure}
    \hfill
    \begin{subfigure}{0.9\columnwidth}
        \centering
        \includegraphics[width=\textwidth]{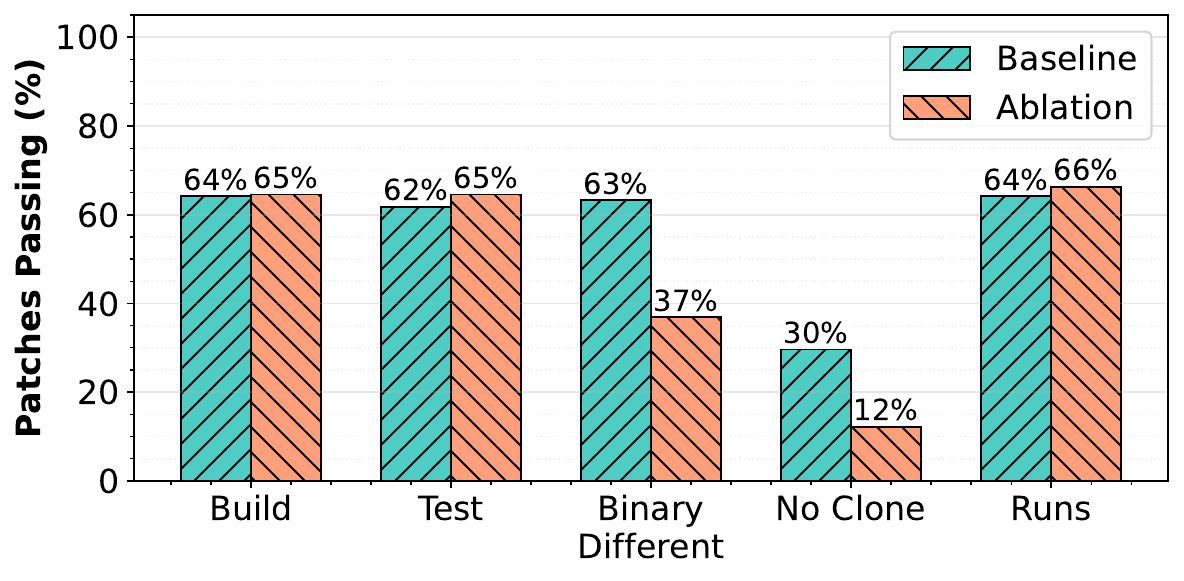}
        \caption{Without the prompt to diversify the implementation we see a drop in patches passing the diversification safeguards.}
        \label{fig:ablation_diversification}
    \end{subfigure}

    \caption{Effects of removing the context and the assignment of diversification from the prompt (baseline) in Listing~\ref{listing:prompt}.}
    \label{fig:ablation}
\end{figure}

\textbf{Ablation study.}
What are the effects of the prompt on the results?
To investigate this, we removed parts of the prompt seen in Listing~\ref{listing:prompt} and showed the results for PBFT.
First, we removed the context (data structure, function signatures) from the prompt.
We can see in Figure~\ref{fig:ablation_context} that the success rate of the functional equivalence safeguards (build, test, runs) drops significantly.
This is explained by the fact that without the context about the codebase the \acp{LLM} hallucinate, e.g., which fields a struct might have.
With these hallucinations the code does not compile, or if it does, functions are not used correctly, which results in failed test and run safeguards.
Quite paradoxically, these hallucinations also explain the increase in patches that pass through the clone safeguard.
The hallucinated code does not exist in the code base, so even though it does not compile, for this test it is a success as it is not a clone.
This also highlights the need for holistic safeguards considering not only diversification but also at functional equivalence.

Secondly, we investigated if we need to nudge the \acp{LLM} to create diversified code.
In Figure~\ref{fig:ablation_diversification} we show the results of removing the call to diversify the code from the prompt.
As expected without the specific constraint for diversified code the \acp{LLM} do not create diversified code.
More patches fail at the binary check and at the clone detection.
This highlights the importance of the prompt on the results.

\begin{figure}[]
    \centering
    \includegraphics[width=1\columnwidth]{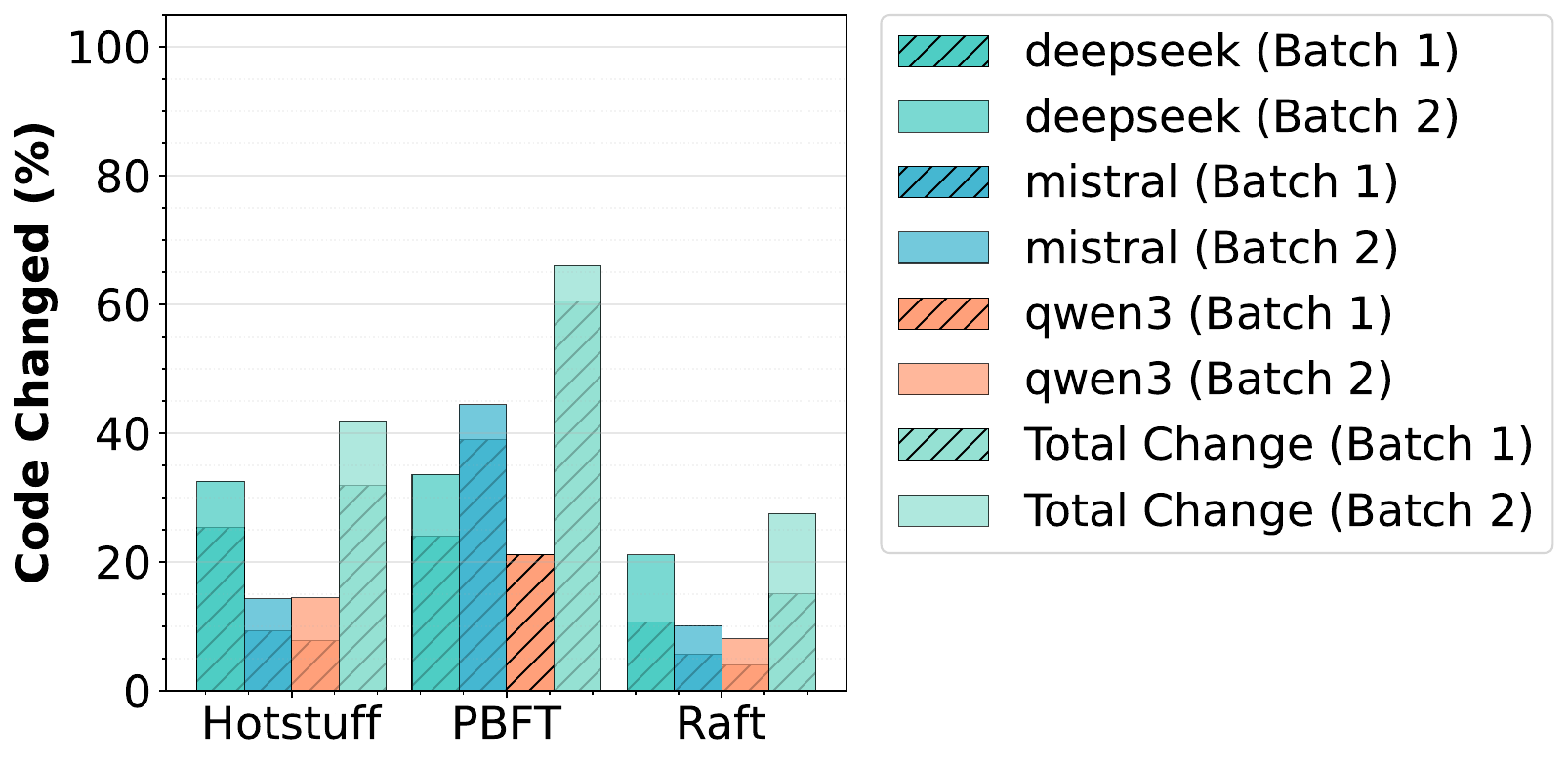}
    \caption{Lines of code replaced from the reference implementation using patches.}
    \label{fig:removed_lines}
\end{figure}

\answer[\qprompt]{Prompts significantly influence the results}{
    We have shown how adding context about the code base reduced hallucinations of \acp{LLM} making them a more reliable tool for diversification.
}

\textbf{Total changes.}
Next, we want to evaluate how much of the original code we can diversify.
To illustrate this, we consider the number of functions that could be diversified and the total lines of code that this change affects.
In Figure~\ref{fig:n-version} we show that running \magic multiple times produces multiple unique patches for functions.
For this, we generated 12 patches for any PBFT function and tested for patches that passed all safeguards whether or not the binary is different between all patches or if there was a clone.
The figure details that we can generate up to \maxUnique unique patches for one function that are not clones and are binary different from each other.
This shows us that running \magic multiple times not only generates a diversified version of the function once, but can also be used to continue generating new patches.
We imagine this to be useful for deployments using techniques such as Proactive Recovery~\cite{castro2002proactive}.
This indicates that we can obtain multiple patches for most functions.

\Cref{fig:removed_lines} shows how many lines of code remain from the original code after applying patches.
We managed to replace up to \locRemovedPercent of the lines of code with just two runs of \magic.
In the experiment, we ran \magic twice to create two batches of patches, for each function we generated 6 patches, two with each \ac{LLM}.
For Raft we saw an increase of 73.8\% of unique patches between batch one and two, 14.7\% for PBFT, and 32.1\% for HotStuff.
The comparably low code change percentage for Raft is explained by the programming style of OpenRaft.
The library is modelled with extremely large match (Rusts switch-case) constructions~\cite{openraft_switch_1, openraft_switch_2, openraft_switch_3}, including nine functions with more than 100 \ac{LoC}.
None of these 100+ \ac{LoC} functions was successfully diversified.

\answer[\qpercent \qmul]{We can diversify large parts of code}{
    We can generate multiple patches for most functions, even with a few runs of \magic.
    Up to \locRemovedPercent of the lines of code could be replaced with validated patches.
}

\begin{figure}[]
    \centering
    \includegraphics[width=1\columnwidth]{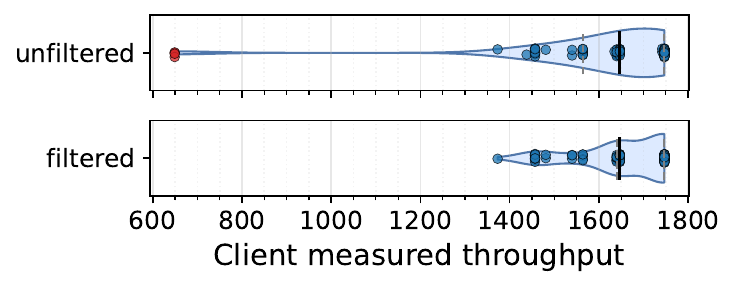}
    \caption{Client measured throughput for diversified Themis with patches passing all previous safeguards.}
    \label{fig:max-diverse-crashes}
\end{figure}

\textbf{Stress test barrier \& fault injection.} With the stress test barrier, we check if individually validated patches still work together once every replica runs a different, fully diversified implementation even when we inject faults (like a leader crash). 
The impact of an injected fault can be timing-dependent: The failure of the replicated system might only occur under specific message orderings \textit{and} only when particular patches interact. 
On a real testbed such faults could behave as Heisenbugs~\cite{gray1986why}: they may appear in one run and never recur. 
This makes it difficult to attribute them to a specific patch or set of patches. 
We therefore run this experiment in Shadow~\cite{rob2022shadow}, a discrete-event network simulator that drives the replica processes with a virtual clock and a seeded, deterministic scheduler. 
Two of its properties are essential to our fault-injection methodology: 

First, Shadow is deterministic: a given configuration and seed reproduce the exact same sequence of events on every run. 
This is what makes our fault-attribution procedure sound. 
We can fix the seed to \emph{reproduce} a failing schedule exactly. 

Second, Shadow lets us inject a deliberate leader crash at an identical point in virtual time across every run and every diversified configuration. 
The view-change path is thus triggered under identical conditions, so the only variable that differs between a failed and a fault-tolerant execution (in which replicas still make progress despite the leader crash) is the set of applied patches and this is precisely what our attribution requires. 
Note, we use Shadow only to \emph{detect} and localize faults for this experiment; later, we measure absolute throughput and latency separately through a system evaluation in a real testbed.

We applied patches from the whole set of patches to the PBFT implementation, 95 patches in total, with different patches for each replica depending on a seed.
We tested four replicas together, shutting the leader down after 20 seconds, and evaluated the throughput of the system.
If the throughput was below a threshold we identified as faulty (e.g., when the system stopped after the view change), then the patches together are identified as faulty.
Keeping the seed for Shadow fixed, we repeated this experiment 100 times, each with a different seed for which patches to apply for each execution.

We initially identified three faulty patches while executing four differently diversified implementations using Shadow.
These were identified by bisecting the set of patches during those executions.
We identified that these functions were not tested so far, accordingly we extended the test corpus for these functions.
This allowed us to retroactively identify these patches as faulty in the test phase.
Next, we reran the Shadow experiment with those three patches filtered out.
The results can be seen in~\Cref{fig:max-diverse-crashes}.
The figure shows the client measured throughput.
Using the extended testing safeguard and the three faulty patches removed, we see three anomalous throughput numbers.
We identified this as three executions where a set of patches caused the view-change to fail.
We used the procedure described in the design to identify 17 potential 3-element sets of patch-replica pairs responsible for the fault, which we then tested again in isolation, resulting in 10 verified subsets and 4 excluded patches.
Rerunning Shadow with a total of 7 (3 individually identified + 4 of the 3-set patches) of 1224 patches from the unfiltered execution removed from the stress test, we see that all 100 executions succeed in the filtered execution.

\textbf{Performance.}
With the set of patches that have passed the stress test, we conducted a performance evaluation with fully diversified replicas in a testbed.
For this we used c6525-25g machines (16-core AMD 7302P at 3.00GHz with 128GB ECC Memory (8x 16 GB 3200MT/s RDIMMs), two 480 GB 6G SATA SSDs and two dual-port Mellanox ConnectX-5 25Gb NICs (PCIe v4.0)) on Cloudlab with 4 replicas and one client, each on their own machine.
For replicas, we used a batch size of 100, with a batch timeout of 10ms, all replicas responding to the client, each with 1000 bytes.
There is one client with configurable concurrent requests, varying request size, and 300 seconds of benchmark time, measuring throughput and latency.

\begin{figure}[]
    \centering
    \includegraphics[width=1\columnwidth]{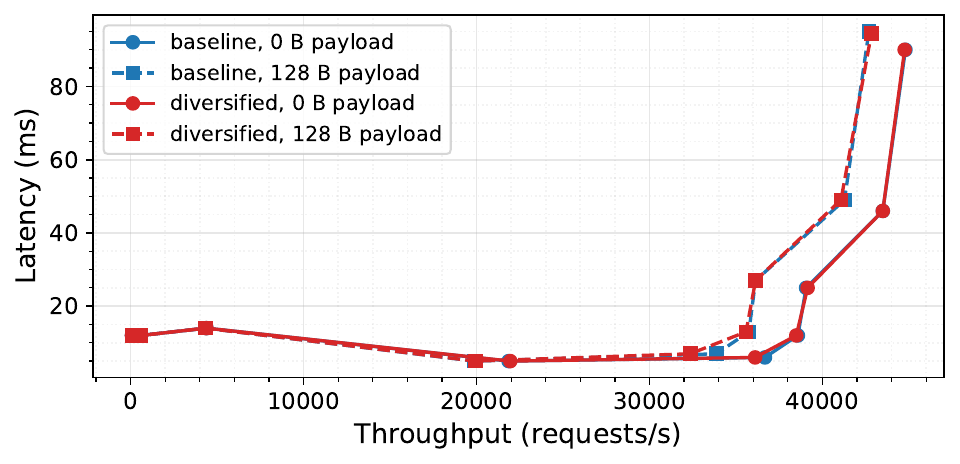}
    \caption{Client measured throughput vs. latency with fully diversified replicas.}
    \label{fig:latency-throughput}
\end{figure}

The results can be seen in~\Cref{fig:latency-throughput}.
For the baseline, we used undiversified Themis, for the diversified measurement, we used 10 different seeds, showing the mean of all measurements.
As can be seen in \Cref{fig:latency-throughput} the difference between the baseline and the diversified execution is minimal.
To be precise, the geometric mean of the absolute differences between the baseline and the diversified execution for each measurement is 0.25\% for throughput and 0.53\% for the latency.
Note on \Cref{fig:latency-throughput}: the first three data points had concurrent request of (1, 8, 64) lower than the batch size (100), resulting batches not filling up and only triggered on the batch timeout (10 ms), resulting in increased latency and low throughput.

\answer[\qstress]{Stress testing identifies remaining faults}{
    Patches cannot be evaluated individually; some issues only occur in cross-interactions between patches.
    With bisecting and common k-set statistics we can identify these without the need for combinatorially many tests.
    Using this, we achieve 100\% success rate during stress tests.
}

\begin{figure}[]
    \centering
    \includegraphics[width=1\columnwidth]{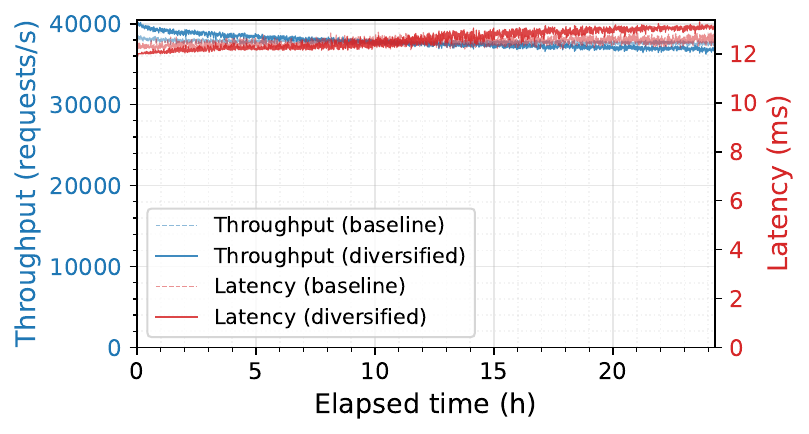}
    \caption{Client measured throughput and latency over a 24 hour stress test. Showing a rolling average of 60 seconds.}
    \label{fig:twenty-four}
\end{figure}

\textbf{Twenty-four-hour test.}
As a final test, we ran Themis PBFT for 24 hours (24.28h = 87400 seconds, to be precise) in a fully diversified configuration on the five c6525-25g machines setup of the stress test, four for the replicas and one for the client.
The results can be seen in~\Cref{fig:twenty-four}.
The plot shows a 60 second rolling average of client measured throughput (min. 32657, median 37773, and max. 41238) and latency (min. 11ms, median 13ms, and max. 15ms).
We observe no fault occurring over the duration of the experiment, processing 3,297,273,515 total requests.
We believe that the slight decrease in performance (both for the baseline and the diversified replicas) over time is caused by thrashing~\cite{bessani2013efficiency, bessani2014trashing}.

\answer[\qstress \qeffect]{Test of time}{
    In a final test running Themis for 24 hours, we did not observe any fault introduced by fully validated patches ordering 3,297,273,515 requests.
}

\textbf{Summary.}
We have shown that \acp{LLM} can diversify up to \locRemovedPercent with two batches of diversification, without observing a single failure in 100 fully diversified executions (\Cref{fig:max-diverse-crashes}).
Even large functions can be diversified (\Cref{fig:loc_bins}), with up to 7 unique new implementations for one function (\Cref{fig:n-version}), and while temperature has little effect (\Cref{fig:llm-temperature}), prompt engineering affects the success rate of diversification (\Cref{fig:ablation}).
Diversification has little effect on performance (\Cref{fig:latency-throughput}: 0.25\% for throughput, 0.53\% for the latency), and runs without any issues in a 24 hour test (\Cref{fig:twenty-four}).


\section{Discussion}
\label{sec:limitations}

\textbf{Bugs that span multiple functions.}
\proj focuses on diversifying code for one function at a time.
This enables us to address bugs at this level (see Section~\ref{sec:proof-of-concept}). 
However, it makes it unlikely that we can fix bugs that span multiple functions.
This is shown in Listing~\ref{listing:request}, where \acp{LLM} were unable to fix another bug we obtained for Themis because of the limitation of having to maintain the function signature.
In the \acf{CoT} of the \ac{LLM} we saw that it correctly identified the problem, but since the solution (as seen by the human fix) required a changed signature, it did not implement the fix.
For issues that span multiple functions, a promising area for future research is the simultaneous generation of code for multiple functions (e.g., connected functions in the call graph).

\begin{listing}[]
    \begin{codebox}
        \begin{minted}[fontsize=\footnotesize, autogobble, linenos=false]{rust}
fn handle_request(
    mut request: Message<Request>,
    mut sender: Sender<RawMessage<ClientTag>>,
    mut recv: Receiver<Message<Response>>,
    peers: usize,
+   faults: usize,
) -> impl Future<Output = Result<Message<Response>>> {
-   let mut quorum = Quorum::new(peers / 3 + 1, peers);
+   let mut quorum = Quorum::new(faults + 1, peers);
// LLM CoT output:
// Actually, the PBFT paper says that the quorum size is f+1
// So if `faults` is f, then quorum size should be faults+1.
// ...
// Since we are not changing the function signature, we must
// use the same quorum parameters. So we will keep
// `Quorum::new(peers / 3 + 1, peers)`.
        \end{minted}
    \end{codebox}
\caption{Human only fix for a wrong quorum in Themis. The LLM noticed the wrong quorum size but was not allowed to fix the bug.}
    \label{listing:request}
\end{listing}

\textbf{Concurrency bugs.}
Concurrency bugs commonly span multiple functions with shared state.
Fortunately, by using Rust, we can use the type system to prevent many concurrency bugs.
Using static analysis tools, we could identify which functions concurrently access shared state and generate code for these functions together.

\textbf{Proactive Recovery.}
Orthogonal to the methodology of how to automate diversification using \acp{LLM} with \magic, we can envision a use case for proactive recovery as proposed by Castro and Liskov~\cite{castro2002proactive}.
In practice, proactively replacing replicas with newly diversified ones could be advantageous to keep the exploitation rate of a  system low, as discussed in the attacker model (§\ref{sec:system_model}).



\section{Conclusion}
\label{sec:conclusion}

If a replicated system runs the same implementation on every replica, then it is only as safe as that implementation. 
Thus, a single common bug can lead to a system failure. 
Diverse protocol implementations can remove a single point of failure, but producing them by hand has been prohibitively expensive. 
In this paper, we propose \magic which tackles this problem: \magic uses several \acp{LLM} to diversify individual functions into versions that are functionally equivalent but representationally and binary-different, by validating each through build-and-test, clone detection, binary comparison, and a stress test that deterministically injects faults.

Our evaluation shows that the overall approach is practical. 
If a patch builds, it will have a high chance of passing every safeguard, and while commercial models perform best, open-weight models alone already provide a reproducible lower bound. 
Apart from this, we found that supplying code-base context can further improve the success rate by reducing hallucination. 
Without developer effort, \magic diversifies up to \locRemovedPercent of a \ac{BFT} codebase with up to \maxUnique distinct variants per function. 
As some faults might only surface in the cross-interaction of patches, we localize and remove them with bisection and common $k$-set statistics rather than combinatorial testing. 
After filtering, we observed no failure across 100 fully diversified executions and in a 24-hour run.

\bibliographystyle{ACM-Reference-Format}
\bibliography{base}

%

\end{document}